\documentclass[12pt]{amsart}
\usepackage[T1]{fontenc}
\usepackage{amssymb}
\usepackage{a4wide}
\usepackage{graphicx}
\usepackage{verbatim}
\usepackage{amsmath}
\usepackage{version}

 \theoremstyle{plain}    
 \newtheorem{thm}{Theorem}[section]
 \numberwithin{equation}{section} 
 \numberwithin{figure}{section} 
 \theoremstyle{plain}
 \newtheorem{prop}[thm]{Proposition} 
 \theoremstyle{plain}    
 \newtheorem{lem}[thm]{Lemma} 
 \theoremstyle{plain}    
 \newtheorem{cor}[thm]{Corollary} 
 \theoremstyle{plain}    
 \newtheorem{conj}[thm]{Conjecture} 
\theoremstyle{definition}
\newtheorem{rem}{Remark} 

\begin{document}
\newcommand{\nwc}{\newcommand}
\nwc{\nwt}{\newtheorem}
\nwt{coro}{Corollary}


\nwc{\mf}{\mathbf} 
\nwc{\blds}{\boldsymbol} 
\nwc{\ml}{\mathcal} 


\nwc{\lam}{\lambda}
\nwc{\del}{\delta}
\nwc{\Del}{\Delta}
\nwc{\Lam}{\Lambda}
\nwc{\elll}{\ell}

\nwc{\IA}{\mathbb{A}} 
\nwc{\IB}{\mathbb{B}} 
\nwc{\IC}{\mathbb{C}} 
\nwc{\ID}{\mathbb{D}} 
\nwc{\IE}{\mathbb{E}} 
\nwc{\IF}{\mathbb{F}} 
\nwc{\IG}{\mathbb{G}} 
\nwc{\IH}{\mathbb{H}} 
\nwc{\IN}{\mathbb{N}} 
\nwc{\IP}{\mathbb{P}} 
\nwc{\IQ}{\mathbb{Q}} 
\nwc{\IR}{\mathbb{R}} 
\nwc{\IS}{\mathbb{S}} 
\nwc{\IT}{\mathbb{T}} 
\nwc{\IZ}{\mathbb{Z}} 
\def\bbbone{{\mathchoice {1\mskip-4mu {\rm{l}}} {1\mskip-4mu {\rm{l}}}
{ 1\mskip-4.5mu {\rm{l}}} { 1\mskip-5mu {\rm{l}}}}}



\nwc{\va}{{\bf a}}
\nwc{\vb}{{\bf b}}
\nwc{\vc}{{\bf c}}
\nwc{\vd}{{\bf d}}
\nwc{\ve}{{\bf e}}
\nwc{\vf}{{\bf f}}
\nwc{\vg}{{\bf g}}
\nwc{\vh}{{\bf h}}
\nwc{\vi}{{\bf i}}
\nwc{\vj}{{\bf j}}
\nwc{\vk}{{\bf k}}
\nwc{\vl}{{\bf l}}
\nwc{\vm}{{\bf m}}
\nwc{\vn}{{\bf n}}
\nwc{\vo}{{\it o}}
\nwc{\vp}{{\bf p}}
\nwc{\vq}{{\bf q}}
\nwc{\vr}{{\bf r}}
\nwc{\vs}{{\bf s}}
\nwc{\vt}{{\bf t}}
\nwc{\vu}{{\bf u}}
\nwc{\vv}{{\bf v}}
\nwc{\vw}{{\bf w}}
\nwc{\vx}{{\bf x}}
\nwc{\vy}{{\bf y}}
\nwc{\vz}{{\bf z}}
\nwc{\bal}{\boldsymbol{\alpha}}
\nwc{\bep}{\boldsymbol{\epsilon}}
\nwc{\bnu}{\boldsymbol{\nu}}
\nwc{\bmu}{\boldsymbol{\mu}}



\nwc{\bk}{\blds{k}}
\def\k{\blds{k}}
\nwc{\bm}{\blds{m}}
\nwc{\bp}{\blds{p}}
\nwc{\bq}{\blds{q}}
\nwc{\bn}{\blds{n}}
\nwc{\bv}{\blds{v}}
\nwc{\bw}{\blds{w}}
\nwc{\bx}{\blds{x}}
\nwc{\bxi}{\blds{\xi}}
\nwc{\by}{\blds{y}}
\nwc{\bz}{\blds{z}}


\nwc{\cA}{\ml{A}}
\nwc{\cB}{\ml{B}}
\nwc{\cC}{\ml{C}}
\nwc{\cD}{\ml{D}}
\nwc{\cE}{\ml{E}}
\nwc{\cF}{\ml{F}}
\nwc{\cG}{\ml{G}}
\nwc{\cH}{\ml{H}}
\nwc{\cI}{\ml{I}}
\nwc{\cJ}{\ml{J}}
\nwc{\cK}{\ml{K}}
\nwc{\cL}{\ml{L}}
\nwc{\cM}{\ml{M}}
\nwc{\cN}{\ml{N}}
\nwc{\cO}{\ml{O}}
\nwc{\cP}{\ml{P}}
\nwc{\cQ}{\ml{Q}}
\nwc{\cR}{\ml{R}}
\nwc{\cS}{\ml{S}}
\nwc{\cT}{\ml{T}}
\nwc{\cU}{\ml{U}}
\nwc{\cV}{\ml{V}}
\nwc{\cW}{\ml{W}}
\nwc{\cX}{\ml{X}}
\nwc{\cY}{\ml{Y}}
\nwc{\cZ}{\ml{Z}}

\nwc{\tA}{\widetilde{A}}
\nwc{\tB}{\widetilde{B}}

\nwc{\To}{\longrightarrow} 

\nwc{\ad}{\rm ad}
\nwc{\eps}{\epsilon}
\nwc{\ep}{\epsilon}
\nwc{\vareps}{\varepsilon}

\def\ep{\epsilon}
\def\tr{{\rm tr}}
\def\Tr{{\rm Tr}}
\def\i{{\rm i}}
\def\mi{{\rm i}}
\def\e{{\rm e}}
\def\sq2{\sqrt{2}}
\def\sqn{\sqrt{N}}
\def\vol{\mathrm{vol}}
\def\defi{\stackrel{\rm def}{=}}
\def\t2{{\mathbb T}^2}
\def\tt2{{\mathbb T}^2}
\def\s2{{\mathbb S}^2}
\def\hn{\mathcal{H}_{N}}
\def\shbar{\sqrt{\hbar}}
\def\S{\mathcal{S}}
\def\A{\mathcal{A}}
\def\N{\mathbb{N}}
\def\T{\mathbb{T}}
\def\R{\mathbb{R}}
\def\Z{\mathbb{Z}}
\def\C{\mathbb{C}}
\def\O{\mathcal{O}}
\def\Sp{\mathcal{S}_+}
\nwc{\rest}{\restriction}
\nwc{\diam}{\operatorname{diam}}
\nwc{\Res}{\operatorname{Res}}
\nwc{\Spec}{\operatorname{Spec}}
\nwc{\Vol}{\operatorname{Vol}}
\nwc{\Op}{\operatorname{Op}}
\nwc{\supp}{\operatorname{supp}}
\nwc{\Span}{\operatorname{span}}

\def\hto0{\xrightarrow{h\to 0}}
\def\htoo{\stackrel{h\to 0}{\longrightarrow}}
\def\rto0{\xrightarrow{r\to 0}}

\providecommand{\abs}[1]{\lvert#1\rvert}
\providecommand{\norm}[1]{\lVert#1\rVert}
\providecommand{\set}[1]{\left\{#1\right\}}

\nwc{\la}{\langle}
\nwc{\ra}{\rangle}
\nwc{\lp}{\left(}
\nwc{\rp}{\right)}

\nwc{\bequ}{\begin{equation}}
\nwc{\ben}{\begin{equation*}}
\nwc{\bea}{\begin{eqnarray}}
\nwc{\bean}{\begin{eqnarray*}}
\nwc{\bit}{\begin{itemize}}
\nwc{\bver}{\begin{verbatim}}

%\nwc{\eal}{\end{align}}
\nwc{\eequ}{\end{equation}}
\nwc{\een}{\end{equation*}}
\nwc{\eea}{\end{eqnarray}}
\nwc{\eean}{\end{eqnarray*}}
\nwc{\eit}{\end{itemize}}
\nwc{\ever}{\end{verbatim}}

\newcommand{\defeq}{\stackrel{\rm{def}}{=}}

\title[Entropy of semiclassical measures]
{Entropy of semiclassical measures of the Walsh-quantized baker's map}

\author[N. Anantharaman]{Nalini Anantharaman}
\author[S. Nonnenmacher]{St\'ephane Nonnenmacher}
\address{Unit\'e de Math\'ematiques Pures et Appliqu\'ees, \'Ecole Normale Sup\'erieure,
6, all\'ee d'Italie, 69364 LYON Cedex 07, France}
\email{nanantha@umpa.ens-lyon.fr}
\address{Service de Physique Th\'eorique, 
CEA/DSM/PhT, Unit\'e de recherche associ\'e CNRS,
CEA/Saclay,
91191 Gif-sur-Yvette, France}
\email{snonnenmacher@cea.fr}

\begin{abstract}We study the baker's map and its Walsh
quantization, as a
toy model of a quantized chaotic system. We focus on localization properties of 
eigenstates, in the semiclassical r\'egime. 
Simple counterexamples show that quantum unique ergodicity
fails for this model. We obtain, however, lower bounds
on the entropies associated with semiclassical measures, as well as on the 
Wehrl entropies of eigenstates.
The central tool of the proofs is an ``entropic uncertainty principle''.
\end{abstract}

\maketitle

\section{Introduction}

In the semiclassical (highly-oscillatory) framework, one can generally express the solution
of the time-dependent Schr\"odinger equation as an $\hbar$-expansion
based on the classical motion. Classical mechanics is then the $0$-th order
approximation to wave mechanics.

However, such expansions are not uniform in time, and generally fail 
to capture
the infinite-time evolution of the quantum
system, or its stationary properties. Unless the system is completely integrable, the 
instabilities of the classical dynamics will
ruin the semiclassical expansion beyond the \emph{Ehrenfest time}, which is of order
$|\log\hbar|$.

Nevertheless, the domain dubbed as ``quantum chaos" expresses the
belief that 
strongly chaotic properties of the classical system
induce certain typical patterns in the stationary properties of the quantum
system, like the statistical properties of the
eigenvalues (the Random Matrix conjecture \cite{bohigas}), or the
\emph{delocalization} of the eigenfunctions over the full accessible phase
space \cite{berry77,voros77}. 


The first rigorous result in this frame of ideas is the ``Quantum Ergodicity Theorem'' 
\cite{Shni74}:
it states that, if the classical system is ergodic on the accessible phase space 
(the energy shell for a Hamiltonian system, respectively the full phase space for an
ergodic symplectic map), then, in the
semiclassical r\'egime, ``almost all'' the eigenstates
become uniformly distributed on that phase space. 
This stands in sharp contrast 
to the case of completely integrable systems, where
eigenstates are known to be localized near well-prescribed
Liouville-Arnold tori, due to a maximal number of invariants of the motion.
``Quantum Ergodicity'' has first been proven for the eigenstates
of the Laplacian on surfaces of negative curvature \cite{CdV85,Zel87}, then for general 
Hamiltonians \cite{HelMarRob87},
ergodic Euclidean billiards \cite{GerLei93,ZelZwo96}, quantized ergodic maps 
\cite{BouzDB96,Zel97} or $C^*$-dynamical systems \cite{Zel96}.

The ``Quantum Unique Ergodicity''
conjecture  goes further in this direction: originally expressed in the
framework of geodesic flows on compact manifolds of negative curvature \cite{RudSar94}, it 
predicts that, for a
strongly chaotic system, \emph{all} the eigenstates should 
be uniformly distributed on the accessible phase space, in the
semiclassical limit.

This conjecture has been tested on a number of models. If the classical system admits
a unique invariant measure, then it boils down to a proof of the quantum-classical
correspondence; Quantum Unique Ergodicity has thus been proven for several families 
of uniquely ergodic maps
on the torus \cite{BouzDB96,MarRud00,Rosenzweig05}.

On the opposite, Anosov systems
admit a vast variety of invariant measures. Applied to these systems, 
the conjecture 
states that quantum mechanics singles out a unique measure out of the set of 
invariant ones.
So far, the conjecture has only be proven for Anosov systems enjoying an 
arithmetic structure, in the form of a commutative algebra of
Hecke operators: this allows to define a preferred eigenbasis of the quantum system, 
namely the joint eigenbasis of all Hecke operators. 
Number theory comes to the rescue of dynamics
to understand these eigenstates \cite{RudSar94,Wol01, BLi03}. E.~Lindenstrauss 
proved the semiclassical 
equidistribution of all Hecke eigenstates of the Laplacian
on compact arithmetic surfaces \cite{Linden};
in that case, the eigenstates of the Laplacian 
are believed to be nondegenerate, which would
make the ``Hecke'' condition unnecessary. 

Studying the quantized automorphisms of the
2-torus (or ``quantum cat maps''), Kurlberg and Rudnick had exhibited such 
a commutative Hecke algebra, and proven that all joint eigenstates
become equidistributed as $\hbar\to 0$ \cite{KurRud00}.
However, the eigenvalues of quantum cat maps 
can be highly degenerate when Planck's constant belongs to a certain sparse 
sequence $(\hbar_k\to 0)$: 
imposing the Hecke condition then
strongly reduces the dimensions of the eigenspaces.
In particular, it was shown in \cite{FNdB03} that, along the same sequence
$(h_k)$, 
certain non Hecke eigenstates
can be partly localized near a classical periodic orbit,
therefore disproving Quantum Unique Ergodicity for the quantum cat maps.
Still, the localized part of the eigenstate
cannot represent more that {\em one half} of 
its total mass \cite{BonDB2,FN04}.
Very recently, Kelmer obtained interesting results about quantized symplectomorphisms of
higher-dimensional tori \cite{Kelmer05}:
if the classical automorphism admits a rational isotropic invariant subspace, he
exhibits a family of Hecke eigenstates (he calls ``superscars''), which are 
fully localized on a dual invariant submanifold.

In the present paper we study another toy model, 
the baker's map defined in
terms of an integer parameter $D\geq 2$ (we will sometimes call this map the $D$-baker).
It is a well-known canonical map on the 2-torus, which is 
uniformly hyperbolic (Anosov) with uniform
Liapounov exponent $\lambda=\log D$. Its Weyl quantization 
\cite{BV,Sar} has been a popular model of ``quantum chaos'' in the last twenty years. 
We will use here a different quantization, based on the Walsh-Fourier
transform \cite{NZ1}: this choice makes the quantum model amenable to 
an analytic treatment.
The map and its quantization
will be described in more detail in Sections~\ref{s:defs}-\ref{s:walsh}. 
The localization in phase
space of an eigenfunction $\psi_\hbar$ will be analyzed using its Walsh-Husimi
measure $WH_{\psi_\hbar}$, which is a probablity measure on the torus, associated
with the state $\psi_\hbar$.
For any sequence of
eigenfunctions $(\psi_\hbar)_{\hbar\to 0}$ of the quantized map, 
one can extract a subsequence
of $\big(WH_{\psi_{\hbar_j}}\big)_{\hbar_j\to 0}$ which weakly converges towards
a probability measure $\mu$. We call such a limit $\mu$ a \textbf{semiclassical measure}.
From the quantum-classical correspondence, $\mu$ is invariant 
through the classical baker's map. Like any Anosov system, the
baker's map admits
plenty of invariant measures: for instance, each periodic orbit carries
an invariant probability measure; we will also describe some (multi)fractal
invariant measures. 

Since the baker's map is ergodic with respect to the
Lebesgue measure, we can easily prove Quantum Ergodicity for the Walsh-quantized map,
stating that the limit measure $\mu$ is ``almost surely'' the Lebesgue measure
(Theorem~\ref{c:QE}). 

Yet, in Section~\ref{sub:particular-eigenstates} we will exhibit some 
examples of semiclassical measures
\emph{different} from the Lebesgue measure, thereby disproving Quantum Unique Ergodicity
for the Walsh-quantized baker. We notice that, as in the case of the quantum cat map,
the presence of partially localized eigenstates 
is accompanied by very high spectral degeneracies. 

Our goal is to characterize 
the possible semiclassical limits $\mu$ among the set of invariant measures.
The tools we will use for this aim are the various {\bf entropies} associated with
invariant measures \cite{KatHas95} (we will recall the definitions of these
entropies).
Our first theorem characterizes the support of $\mu$.
\begin{thm} 
\label{HTOP} 
Let $\mu$ be a semiclassical measure of the Walsh quantized
$D$-baker, and $\supp \mu$ its support. 
The topological entropy of that support must satisfy
$$
h_{top}(\supp \mu)\geq \frac{\log D}{2}=\frac{\lambda}{2}\,.
$$
\end{thm}
The theorem implies, in particular, that the measure $\mu$ cannot be entirely 
concentrated on periodic orbits (for any periodic orbit $\cO$, $h_{top}({\cO})=0$);
it still allows its support to be 
thinner than the full torus ($h_{top}(\t2)=\log D$). 
This theorem was proved in \cite{An}
for the eigenstates of the Laplacian on compact Riemannian manifolds
with Anosov geodesic flows. 
The proof of Theorem~\ref{HTOP} presented below uses the same strategy, 
but is made much
shorter by the simplicity of the particular model (see Section~\ref{s:HTOP}). 
In fact, we present Theorem~\ref{HTOP}
mostly for pedagogical reasons, since we can prove a stronger result: 
\begin{thm} 
\label{HMET} 
Let $\mu$ be a semiclassical measure of the Walsh quantized
$D$-baker. Then its Kolmogorov-Sinai entropy satisfies
$$
h_{KS}(\mu)\geq \frac{\log D}{2}=\frac{\lambda}{2}\,.
$$
\end{thm}
Theorem \ref{HMET} is stronger than \ref{HTOP}, because of the Ruelle-Pesin
inequality, $h_{KS}(\mu)\leq\nolinebreak h_{top}(\supp \mu)$ \cite[Theorem 4.5.3]{KatHas95}. 
For instance,
the counterexamples to Quantum Unique Ergodicity constructed in \cite{FNdB03} for 
the quantum cat map
satisfy $h_{top}(\supp \mu)=\lambda$ (the support of $\mu$ is the full torus), 
but $h_{KS}(\mu)=\frac{\lambda}{2}$, showing that the above lower bound
is sharp in that case 
(here, $\lambda$ is the positive Liapounov exponent for the cat map).
In the case of the Walsh-baker's map, 
we will exhibit examples of semiclassical measures $\mu$ which saturate 
the lower bound $\frac{\log D}{2}$ for both the metric entropy $h_{KS}(\mu)$
and the topological entropy $h_{top}(\supp\mu)$ 
(see Section~\ref{sub:particular-eigenstates}). 
The lower bound of Theorem~\ref{HMET} is somehow half-way between 
a completely localized measure ($h_{KS}(\delta_{\cO})=0$ if $\delta_{\cO}$
is the invariant measure carried on a periodic orbit $\cO$) and
the equidistribution ($h_{KS}(\mu_{Leb})=\log D$). 

One can decompose any semiclassical measure into its pure point, singular continuous
and Lebesgue parts 
\begin{equation}\label{e:decompo1}
\mu=\beta_{pp}\mu_{pp}+\beta_{sc}\mu_{sc}+\beta_{Leb}\mu_{Leb},\quad 
\text{with}\ \beta_*\geq 0,
\quad \beta_{pp}+\beta_{sc}+\beta_{Leb}=1\,.
\end{equation} 
Because the functional $h_{KS}$ is affine, Theorem~\ref{HMET} straightforwardly implies 
the inequality 
$\beta_{pp}\leq \beta_{sc}+\beta_{Leb}$.
Actually, one can also adapt the
methods of \cite{FN04} to the Walsh-baker, and obtain a sharper inequality
between these weights:
\begin{thm}\label{t:loc-erg}
Let $\mu$ be a semiclassical measure of the Walsh quantized
$D$-baker. The weights appearing in the decomposition \eqref{e:decompo1} must satisfy:
$$
\beta_{pp}\leq \beta_{Leb}\,.
$$
\end{thm}
In \cite{FN04}, the analogous result had raised a question on
the existence of semiclassical measures of purely singular continuous
nature, in the case of the quantum cat map. 
For the Walsh quantized baker, we answer this
question by the affirmative, by 
constructing explicit examples of such semiclassical measures, with
simple self-similarity properties
(see Section~\ref{sub:particular-eigenstates}).

In the course of the proof of Theorem~\ref{HMET}, we obtain a lower bound
for the \emph{Walsh-Wehrl entropies} associated with the individual eigenstates
(these entropies are defined in Section~\ref{s:Wehrl}). 
The ``standard'' Wehrl entropy \cite{Wehrl} has been used
to characterize the localization of eigenstates in ``quantum chaotic'' systems
\cite{Zycz90,NV98}. For the present model, the Walsh-Wehrl entropies of any
eigenstate are equal to its \emph{Shannon entropy},
another indicator of localization \cite{Izra90}. 
\begin{thm}\label{t:Wehrl}
The Wehrl and Shannon entropies of any eigenstate $\psi_\hbar$ of the Walsh quantized
baker are bounded from below as follows:
$$
h_{Wehrl}(\psi_\hbar)=h_{Shannon}(\psi_\hbar)\geq \frac{|\log 2\pi \hbar|}{2}\,.
$$
\end{thm}
Once more, this lower bound is situated ``half-way'' between the case of maximal
localization ($h_{Wehrl}=0$)
and maximal equidistribution ($h_{Wehrl}=|\log 2\pi\hbar|$). 
A ``typical'' state $\psi_\hbar$, drawn from one of the ensemble of Gaussian random states
described in \cite[Section 5.1]{NV98}, will have a Wehrl entropy of order
$h_{Wehrl}(\psi_\hbar)=
|\log 2\pi\hbar|- C\pm \hbar^{1/2}\,|\log\hbar|$, where the last term denotes the 
standard deviation (the constant $C=1-\gamma_{\rm Euler}$ was first 
derived in \cite{Zycz90}). The lower bound $\frac{|\log 2\pi \hbar|}{2}$ is far 
outside this ``typical interval''. We can construct
eigenstates of the Walsh-baker which saturate this lower bound: 
they are quite different from ``typical'' states.

The proof of the above theorem
relies on an ``Entropic
Uncertainty Principle'' \cite{Kraus87,MaaUff}, which is a variation around the 
Heisenberg Uncertainty 
Principle. It gives some 
consistency to the belief that the Uncertainty
Principle (the central property of quantum mechanics), 
combined with the mixing properties of the Anosov dynamics, leads to some degree
of delocalization of the eigenfunctions. 

Another essential ingredient of the proof is the control of the quantum evolution
up to the Ehrenfest time $\frac{|\log \hbar|}{\lambda}$, which is the
time where the quantum-classical correspondence breaks down.
For the Walsh-baker, this evolution can
be described in a simple algebraic way, without any small remainders, which makes
the analysis particularly simple.

In a forthcoming paper we plan to generalize Theorem \ref{HMET} along the
following lines. 
Our aim is to deal with arbitrary Anosov canonical maps on a 
compact symplectic manifold, 
respectively arbitrary Anosov Hamiltonian
flows on some compact energy shell. Quantizing such systems \`a la Weyl 
and studying their eigenstates in the semiclassical limit,
we conjecture the following lower bound 
for the semiclassical measures $\mu$:
\begin{conj}
Let $\mu$ be a semiclassical measure for an Anosov canonical map 
(resp. Hamiltonian flow) on a compact symplectic manifold (resp. a compact
energy shell) $M$. 
Then its Kolmogorov-Sinai entropy should satisfy
$$
h_{KS}(\mu)\geq \frac{1}{2}\int_{M} |\log J^u(x)|\,d\mu(x)\,,
$$
where $J^u(x)$ is the unstable Jacobian \cite{KatHas95} of the system 
at the point $x$.
\end{conj}
In the case of an Anosov geodesic flow, this lower bound is close to 
the one proven by the first author for $h_{top}(\supp\mu)$
\cite{An}.
For a quantized hyperbolic symplectomorphism
of $\T^{2d}$, this lower bound takes the value
$\frac{1}{2}\sum_{|\lambda_j|>1}\log|\lambda_j|$, where one sums over the 
expanding eigenvalues of the classical map.
The ``superscars'' constructed
in \cite{Kelmer05} do indeed satisfy this lower bound.
The proof of that conjecture will necessarily be
more technical than in the present paper, due to the presence of
small remainders, and also the more complicated nonlinear classical dynamics. 

Let us now outline the structure of the paper.
In Section~\ref{s:defs} we describe the model of the classical baker's map.
Its Walsh quantization is presented in Section~\ref{s:walsh}, and some of
its properties are analyzed. Some particular eigenstates with interesting 
localization properties are exhibited in Section~\ref{sub:particular-eigenstates}.
In Section~\ref{s:HMET} we prove Theorems~\ref{HMET} and \ref{t:Wehrl} using the Entropic
Uncertainty Principle. 
Section~\ref{s:HTOP} is 
devoted to the proof of Theorem \ref{HTOP}, using the strategy of \cite{An}.
Finally, in Section~\ref{s:loc-erg} we sketch the proof of Theorem~\ref{t:loc-erg},
adapted from \cite{FN04}. 

\section{The baker's map and its symbolic dynamics.}
\label{s:defs}

\subsection{The baker's map on the torus}
The phase space we consider is the 2-dimensional torus 
$\t2=(\IR/\IZ)^2\equiv [0,1)\times [0,1)$, 
with position (horizontal) and  momentum (vertical) coordinates $x=(q,p)$. 
We select some integer $D>1$, and define the $D$-baker's map $B$ as follows:
\begin{equation}\label{e:baker}
\forall (q,p)\in \t2,\qquad B(q,p)=\big( Dq\mbox{ mod }1,\; 
\frac{p+\lfloor Dq \rfloor}D\big)\in\t2\,.
\end{equation}
Here $\lfloor x \rfloor$ denotes the largest integer smaller or equal to $x\in\IR$.

This map is invertible on $\t2$,  piecewise affine
with discontinuities along the segments
$\set{ p=0} $ and
$\set{q=j/D}$, $j=0,\ldots,D-1$.
In Fig.~\ref{fig:baker} we schematically represent the map in the case $D=3$.
\begin{figure}
\begin{center}
\includegraphics[angle=0,width=14.0cm]{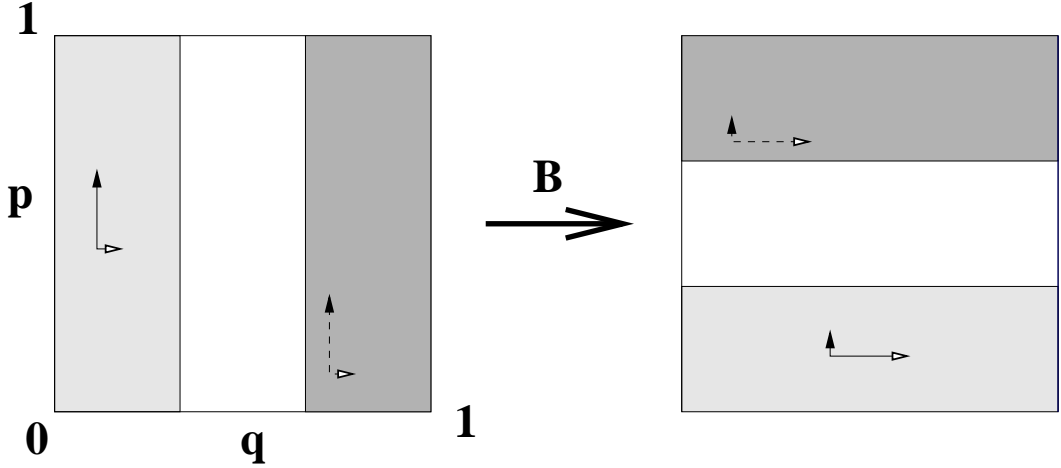}
\caption{Schematic representation of the baker's map \eqref{e:baker} for $D=3$. 
The arrows show the vertical contraction and horizontal dilation.}
\label{fig:baker}
\end{center}
\end{figure}
The map preserves the symplectic form $dp\wedge dq$. 
It is uniformly hyperbolic, with constant Liapounov exponent
$\lambda=\log D$. The stable (resp. unstable) directions are the
vertical (resp. horizontal) directions. 

\subsection{Symbolic dynamics}
The map $B$ can be easily expressed in terms of the $D$-nary
representation of the coordinates $(q,p)$. Indeed, let us
represent the position $q\in[0,1)$ and momentum $p\in[0,1)$ of any
point $x=(q,p)\in\t2$ through their $D$-nary sequences
$$
q=0.\ep_1\ep_2\ldots,\qquad p=0.\ep'_1\ep'_2\ldots\,,\quad\text{where the ``symbols''}\quad
\ep_{i},\, \ep'_{i}\in\set{0,\ldots,D-1}\equiv \IZ_D\, .
$$
We then associate with 
$x=(q,p)$ the following bi-infinite sequence
$$
x\equiv\ldots\ep'_{2}\ep'_{1}\,\cdot\,\ep_{1}\ep_{2}\ldots.
$$
Symbolic sequences will be shortly denoted by
$\bep=\ep_1\ep_2\ldots$, without precising their lengths (either finite or infinite), 
and from there $x\equiv \bep'\cdot\bep$.

More formally, we call $\Sigma_+=\set{0,\ldots,D-1}^{\IN_*}$ the set of 
one-sided infinite sequences, 
and $\Sigma=\Sigma_+\times\Sigma_+$, 
the set of two-sided infinite sequences. The $D$-nary decomposition then generates a map
\begin{align*}
J:\Sigma &\To [0, 1)\times [0,1)\\
\bep'\cdot\bep&\longmapsto x=(0.\bep,\, 0.\bep')\,.
\end{align*} 
The map $J$ is one-to-one except on a denumerable set where it is two-to-one 
(for instance,
$\ldots 00\cdot 100\ldots$ is sent to the same point as $\ldots 11\cdot 011\ldots$).
Let us equip $\Sigma$ with the distance
\begin{equation}\label{e:metrics}
d_\Sigma(\bep'\cdot\bep,\, \bal'\cdot \bal)= \max(D^{-n_0'},\,D^{-n_0})\,,
\end{equation}
where $n_0=\min\set{n\geq 0\,:\,\ep_{n+1}\neq\alpha_{n+1}}$ and similarly for $n_0'$.
The map $J$ is Lipschitz-continuous with respect to this distance. 

$J$ gives a semiconjugacy
between, on one side, the action of $B$ on the torus, on the other side, the simple
{\em shift} on $\Sigma$:
\begin{equation}
\label{eq:baker-shift}
B\big(J( \bep'\cdot\bep)\big)=
J(\ldots\ep'_{2}\ep'_{1}\ep_{1}\,\cdot\,\ep_{2}\ep_{3}\ldots)\,.
\end{equation}
This is a very simple example of symbolic coding of a dynamical system.
The action of $B$ on $\Sigma$ is {\em Lipschitz-continuous}, as opposed to
its discontinuous action on $\t2$ equipped with its standard topology.
As long as we are only interested in characterizing the entropies of invariant measures, 
it is harmless to identify the two systems.
In the following discussion we will go back and forth between the two representations.

\subsection{Topological and metric entropies}\label{sub:entropy}
Let $(X,d)$ be a compact metric space, and $T:X\to X$
a continuous map. In this section, we give the definitions and some properties of the 
topological and metric entropies associated with the map $T$ on $X$. 
We then consider the particular case of the map $B$, seen as the shift acting
on $\Sigma$.

\subsubsection{Topological entropy}  
The topological entropy of the dynamical system $(X, T)$
is defined as follows: for any $n>0$, define the distance 
$$
d_n^T(x, y)\defeq\max_{m=0,...,n}d(T^m x, T^m y)\,.
$$
For any $r>0$, let $N_T(r,n)$ be the minimal cardinal
of a covering of $X$ by balls of radius $r$ for the distance $d_n^T$. 
Then the topological entropy of the set $X$ with respect to the map $T$ is defined as
$$
h_{top}(X, T)\defi \lim_{r \to 0}\limsup_{n\to\infty}\frac1n \log N_T(r,n)\,.
$$
In many cases, it is not necessary to let $r\to 0$: there exists $r_0>0$ such
that, for any $0<r\leq r_0$, the topological entropy 
is equal to 
$\limsup_{n\to \infty}\frac1n \log N_T(r,n)$.

In the case $X=\Sigma$ (equipped with the metrics $d_\Sigma$ given in \eqref{e:metrics}), 
the topological entropy can be expressed using {\em cylinder sets}. 
Given two sequences $\bep$, $\bep'$ of {\em finite} lengths $|\bep|=n$, $|\bep'|=n'$,
we define the cylinder set $[\bep' \cdot \bep]\subset \Sigma$ as the set
of sequences starting with $\bep$ on the right side and 
with $\bep'$ on the left side.
If $n=n'$, it is a ball of radius $D^{-n}$ for the distance $d_\Sigma$.
The image of $[\bep' \cdot \bep]$ on the torus is the rectangle
$$
J([\bep' \cdot \bep])
=[\frac{j}{D^n},\frac{j+1}{D^n}]\times [\frac{j'}{D^{n'}},\frac{j'+1}{D^{n'}}]\,,
\quad\text{where}\quad\frac{j}{D^n}=0.\ep_1\cdots\ep_{n},\ 
\frac{j'}{D^{n'}}=0.\ep'_1\cdots\ep'_{n'}\,.
$$ 
In the following we will often identify cylinders and rectangles.

Since we are interested in the action of the shift,
we can focus our attention to one-sided cylinder sets, of the form 
$[\cdot \bep]$, corresponding
on the torus to ``vertical'' rectangles $[\frac{j}{D^n}, \frac{j+1}{D^n}]\times [0, 1]$.
The set of cylinders $[\cdot\bep]$ of length $n=|\bep|$ will be called $\Sigma_n$.

\bigskip

Let now $F$ be a closed subset of $\Sigma$, invariant under the action of $B$. Call
$N_B(n, F)$ the minimal number of cylinder sets $[\cdot \bep]$ of length $n$ necessary
to cover $F$. The topological entropy $h_{top}(F, B)$, also denoted by
$h_{top}(F)$, is then given by
\begin{equation} \label{e:topo_entropy_F}
h_{top}(F)=\limsup_{n\to\infty} \frac1n \log N_B(n, F)\,.
\end{equation}

\vspace{.2cm}
{\em Examples.} If $F=\cO$ is a periodic orbit, we find $h_{top}(\cO)=0$.
If $F=\t2$, we find $h_{top}(\t2)=\log D$. It is also useful to note that,
if $F$ and $G$ are two closed invariant subsets, then $h_{top}(F\cup G)=\max(
h_{top}(F), h_{top}(G)).$

\subsubsection{Metric entropy} \label{s:metric}
Going back to the general framework, we consider a $T$-invariant probability
measure $\mu$ on the metric space $X$. 

If $\cP=(P_1,...,P_n)$ is a finite measurable partition of $X$ 
(meaning that $X$ is the disjoint
union of the $P_i$s), we define the entropy of the measure $\mu$ with respect to the partition
$\cP$ by
\begin{equation}\label{e:def-entropy}
h_{\cP}(\mu)=-\sum_i \mu(P_i)\log\mu(P_i)\,.
\end{equation}
For $\cP=(P_1,...,P_n)$ and $\cQ=(Q_1,...,Q_m)$ any two partitions of $X$, 
we can define a new partition
$\cP\vee \cQ$ as the partition composed of the sets $P_i\cap Q_j$. 
The entropy has the following
{\em subadditivity property}:
\begin{equation}\label{e:subadd}
h_{\cP\vee \cQ}(\mu)\leq h_{\cP}(\mu)+h_{\cQ}(\mu)\,.
\end{equation}
We may now use the map $T$ to refine a given partition $\cP$:
for any $n\geq 1$ we define the partition
$$
\cP^{(n)}=\cP\vee T^{-1}\cP\vee...\vee T^{-(n-1)}\cP\,.
$$ 
By the subadditivity property, they satisfy
$$
h_{\cP^{(n+m)}}(\mu)\leq h_{\cP^{(n)}}(\mu)+h_{T^{-n}\cP^{(m)}}(\mu)\,.
$$
If the measure $\mu$ is $T$-invariant, $h_{T^{-n}\cP^{(m)}}(\mu)=h_{\cP^{(m)}}(\mu)$.  
The subadditivity of the sequence $\big(h_{\cP^{(n)}}(\mu)\big)_{n\geq 1}$ implies the
existence of the limit:
\begin{equation}\label{def:HMET}
\lim_{n\to\infty}\frac1n\, h_{\cP^{(n)}}(\mu)= \inf_{n\geq 1}\frac1n\, h_{\cP^{(n)}}(\mu)
\defeq h_{\cP}(\mu, T)\,.
\end{equation}
This number $h_{\cP}(\mu, T)$ is the entropy of the measure $\mu$
for the action of $T$, with respect to the partition $\cP$.
The Kolmogorov-Sinai entropy of the triplet $(X, T, \mu)$, denoted by
$h_{KS}(\mu, T)$, is the supremum
of $h_{\cP}(\mu, T)$ over all finite measurable partitions $\cP$. 

\subsubsection{Generating partition for the baker's map}\label{s:generating}
In the case we will be interested in, namely the shift $B$ acting on $\Sigma$,
this supremum is reached if we start from the partition $\cP$ made of the
cylinder sets of length one, that is of the form $[.\ep_1]$ for $\ep_1\in\IZ_D$.
Each such cylinder is mapped on the torus into a vertical rectangles 
$[\frac{\ep_1}{D},\frac{\ep_1+1}{D}]\times [0, 1)$.
Obviously, the refined partition $\cP^{(n)}$ is made of the cylinder sets  
$[.\bep]$ of length $n$, 
representing vertical rectangles 
$[\frac{j}{D^n}, \frac{j+1}{D^n}]\times [0, 1)$.
For any $B$-invariant measure $\mu$ on $\t2$, the metric entropy 
$h_{KS}(\mu, B)=h_{KS}(\mu)$ is given by
\begin{equation}
h_{KS}(\mu)=\inf_{n\geq 1} \frac1n\, h_{\cP^{(n)}}(\mu)=
\lim_{n\to\infty} \frac1n\, h_{\cP^{(n)}}(\mu)\,.
\end{equation}
\vspace{.2cm}

{\em Examples.} If $\mu=\delta_{\cO}$ is an invariant measure carried
on a periodic orbit, we find $h_{KS}(\delta_{\cO})=0$. Another class
of interesting examples are Bernoulli measures~: given some probability weights 
$p_0,...,p_{D-1}$ ($p_\ep\geq 0$, $\sum_{\ep} p_\ep=1$), the infinite product
measure $\mu_{Ber}=\big(\sum_{\ep=0}^{D-1}p_\ep\delta_\ep \big)^{\otimes \Z}$ on $\Sigma$
is invariant under the shift. On $\t2$, it gives a $B$-invariant
probability measure, with simple self-similarity properties. Its Kolmogorov-Sinai entropy is $h_{KS}(\mu_{Ber})=-\sum_{\ep}p_\ep\log p_\ep$. 
The Lebesgue measure corresponds to the case $p_\ep\equiv D^{-1}$ and has maximal entropy,
$h_{KS}(\mu_{Leb})=\log D$. It is also useful to know that the functional
$h_{KS}$ is affine on the convex set of invariant probability measures.

Let us now describe the quantum framework we will be working with.

\section{Walsh quantization of the baker's map}\label{s:walsh}

\subsection{Weyl quantization of the 2-torus}
The usual way to ``quantize'' the torus phase space $\t2$ consists in periodizing
quantum states $\psi\in \cS'(\IR)$ in both position and momentum; the resulting
vector space $\hn$ is nontrivial if and only if Planck's constant 
$\hbar=(2\pi N)^{-1}$, $N\in\IN$, in which
case it has dimension $N$. An orthonormal basis of $\hn$ is given by the ``position
eigenstates'' $\set{ \bq_{j},\; j=0,\ldots,N-1}$ localized at positions $q_j=j/N$.
The ``momentum eigenstates'' are obtained from the latter by applying the inverse of
the Discrete Fourier Transform $\cF_N$,
\begin{equation}\label{e:FT}
(\cF_N)_{jk}=\frac1{\sqrt{N}}\,e^{-2 i \pi kj/N}\,,\quad j,k=0,\ldots,N-1\,.
\end{equation}
This Fourier transform was the basic ingredient used by Balazs and Voros
to quantize the baker's map \cite{BV,Sar}.
Precisely, in the case where $N$ is a multiple of $D$, the (Weyl) quantum baker is
defined as the following  unitary matrix in the position basis:
\begin{equation}\label{e:Weyl-baker}
B_{N}^{BV}=\cF_{N}^{-1}
\begin{pmatrix}
\cF_{N/D} & 0 & 0\\
0 & \ddots & 0\\
0 & 0 & \cF_{N/D}\end{pmatrix}
\end{equation}
These matrices have been studied in detail \cite{SarVor}, but little rigorous is known
about their spectrum. They suffer from diffraction effects due to the
classical discontinuities of $B$ 
(the Egorov property is slightly problematic, but still allows one
to prove Quantum Ergodicity \cite{DENW}). It was recently observed \cite{arul04}
that some eigenstates of the 2-baker in the case
$N=2^k$, ($k\in\IN$) have an interesting multifractal structure in phase space. 
These eigenstates were analyzed using the Walsh-Hadamard transform.

\subsection{Walsh quantum kinematics}
In the present work, we will use the Walsh transform as a building block to quantize
the baker's map. As we will see, the resulting 
Walsh quantization of $B$ respects its $D$-nary
coding, and allows for an exact spectral analysis. 
It has already been used in \cite{NZ1} in the case of ``open'' baker's maps.

Before quantizing the map $B$ itself, we must first describe the Walsh quantum setting
on the 2-dimensional torus, obtained by replacing the usual Fourier transform 
by the Walsh-Fourier transform. The latter was originally defined in the framework
of signal processing \cite{lifermann}. More recently, it has been used as a toy model in
several problems of harmonic analysis (see e.g. the introduction to the ``Walsh phase
space'' in \cite{ThielePhD}).

\subsubsection{Walsh transform}
We will use a Walsh transform {\em adapted} to the
$D$-baker \eqref{e:baker}.
The values of Planck's
constant we will be considering are of the form 
$\set{\hbar=\hbar_k=(2\pi D^{k})^{-1},\ k\in\IN}$,
so the semiclassical limit reads 
$k\to\infty$. 
The quantum Hilbert space is then isomorphic to $\IC^D\otimes\cdots\otimes \IC^D$
(with $k$ factors). More precisely, if we call $\set{e_0,\ldots,e_{D-1}}$ an
orthonormal basis of $\IC^D$, and identify each index $j\in\set{0,\ldots,D^k-1}$
with its $D$-nary expansion $j\equiv \ep_1\cdots\ep_k$, then the isomorphism 
$\cH_{D^k}\simeq (\IC^D)^{\otimes k}$ is
realized through the orthonormal basis of position eigenstates:
\begin{equation}
\bq_{j}=e_{\ep_1}\otimes e_{\ep_2}\otimes\cdots\otimes e_{\ep_k}\,.
\end{equation}  
Each factor space $\IC^D$ is called a ``quantum $D$it'', or qu$D$it, in the 
quantum computing framework. We see that each qu$D$it is associated with a
particular position {\em scale}.

The Walsh transform on $\cH_{D^k}$, which
we denote by $W_{D^k}$, is a simplification of the Fourier transform $\cF_{D^k}$.
It can be defined in terms of the $D$-dimensional
Fourier transform $\cF_{D}$ (see \eqref{e:FT}) through its
action on tensor product states
\begin{equation}\label{eq:walsh-transform}
W_{D^k}(v^{(1)}\otimes\ldots\otimes v^{(k)})=
\cF_{D}v^{(k)}\otimes\cF_{D}v^{(k-1)}\otimes\ldots\otimes\cF_{D}v^{(1)},\qquad
v^{(i)}\in\IC^D,\ i=1,\ldots,k\,.
\end{equation}
The image of position eigenstates through $W_{D^k}^{*}$ yields
the orthonormal basis of {\bf momentum eigenstates}. To each 
momentum $p_{l}=l/D^{k}=0.\ep'_{1}\ldots\ep'_{k}$, $l=0,\ldots,D^k-1$
is associated the state
$$
\bp_{l}=\sum_{j=0}^{D^k-1}\big(W_{D^k}^*\big)_{lj} \bq_j= 
\cF_{D}^{*}e_{\ep'_{k}}\otimes\cF_{D}^{*}e_{\ep'_{k-1}}\otimes
\cdots\otimes\cF_{D}^{*}e_{\ep'_{1}}\,.
$$
Therefore, each qu$D$it also corresponds to a particular momentum scale 
(in reverse order with respect to its corresponding position scale).

From now on, we will often
omit the subscript $D$ on the Fourier transform, and
simply write $\cF=\cF_D$.

\subsubsection{Quantum rectangles and Walsh coherent states}\label{s:CS}
Given any integer $0\leq\ell\leq k$, two sequences 
$\bep=\ep_{1}\ldots\ep_{\ell}\in\Sigma_{\ell}$, 
$\bep'=\ep'_{1}\ldots\ep'_{k-\ell}\in\Sigma_{k-\ell}$ define a 
rectangle $[\bep'\cdot\bep]$ of area $\Delta q\Delta p=D^{-k}=h_k$: for this
reason, we
call it a \emph{quantum rectangle} (in the time-frequency framework \cite{ThielePhD}, such 
rectangles are called {\em tiles}).
To this rectangle we associate the {\em Walsh coherent state}
$|\bep'\cdot\bep'\ra$ defined as follows:
\begin{equation}\label{e:CS}
|\bep'\cdot\bep\ra\defeq 
e_{\ep_{1}}\otimes e_{\ep_{2}}\otimes\ldots 
e_{\ep_{\ell}}\otimes\cF^{*}e_{\ep'_{k-\ell}}\otimes\ldots\otimes\cF^{*}e_{\ep'_{1}}\,.
\end{equation}
For each choice of $\ell$, $0\leq \ell\leq k$, we consider the family of 
quantum rectangles
\begin{equation}\label{e:q.rect.}
\cR^{k,\ell}\defi\set{[\bep'\cdot\bep]\ :\ \bep\in\Sigma_\ell,\ \bep'\in\Sigma_{k-\ell}}\,.
\end{equation}
The corresponding family of coherent states 
$\set{|\bep'\cdot\bep\ra\,:\,[\bep'\cdot\bep]\in \cR^{k,\ell}}$ then
forms an {\em orthonormal} basis of $\cH_{D^k}$, which we will call the $\ell$-basis, or
basis of $\ell$-coherent states. The state $|\bep'\cdot\bep\ra$ is {\em strictly localized} in 
the corresponding rectangle $[\bep'\cdot\bep]$, in the following sense:
$$
\forall j\equiv\alpha_1\ldots\alpha_k,\quad \begin{cases}
|\la \bq_j|\bep'\cdot\bep\ra|&=D^{-\ell/2}\qquad \text{if}\ \alpha_1=\ep_1,\ldots,
\alpha_\ell=\ep_\ell,
\qquad 0\ \text{otherwise}\\
|\la \bp_j|\bep'\cdot\bep\ra|&=D^{-(k-\ell)/2}\quad \text{if}\ \alpha_1=\ep'_1,\ldots,
\alpha_{k-\ell}=\ep'_{k-\ell},\qquad 0\ \text{otherwise}\,.
\end{cases}
$$
This property of strict localization in both position and momentum is the main
reason why Walsh harmonic analysis is easier to manipulate than the usual Fourier analysis
(where such a localization is impossible). Obviously, for $\ell=k$ (resp. $\ell=0$)
we recover the position (resp. momentum) eigenbasis.

Each $\ell$-basis provides a Walsh-Husimi
representation of $\psi\in\cH_{D^{k}}$: it is the
non-negative function $WH^{k,\ell}_{\psi}$ on $\t2$, 
constant inside each rectangle $[\bep'\cdot\bep]\in \cR^{k,\ell}$, where
it takes the value:
\begin{equation}\label{eq:Walsh-Husimi}
WH^{k,\ell}_{\psi}(x)\defi D^{k}\,\left|\la\psi|\bep'\cdot\bep\ra\right|^{2},\qquad 
x\in[\bep'\cdot\bep]\,.
\end{equation}
The standard (``Gaussian'') Husimi function of a state $\psi$ contains all the information about
that state (apart from a nonphysical phase prefactor) \cite{LebVor90}. 
On the opposite, the Walsh-Husimi function 
$WH^{k,\ell}_{\psi}$ only 
contains ``half'' the information on $\psi$ (namely, the moduli of the
components of $\psi$ in the $\ell$-basis). This important difference
will not bother us in the following.

In the case of a tensor-product state $\psi=v^{(1)}\otimes v^{(2)}\otimes\cdots\otimes v^{(k)}$
(each $v^{(i)}\in\IC^{D}$) relevant in Section~\ref{s:tensor-prod}, we have~:
$$
WH^{k,\ell}_{\psi}(x)=D^k\big|v_{\ep_1}^{(1)}\big|^2\ldots 
\big|v_{\ep_{\ell}}^{(\ell)}\big|^{2}\: 
\big|(\cF v^{(k)})_{\ep'_1}\big|^2\ldots 
\big|(\cF v^{(\ell+1)})_{\ep'_{k-\ell}}\big|^2\,,
\qquad x\in[\bep'\cdot\bep]\,.
$$
If $\psi$ is normalized,  
$WH^{k,\ell}_{\psi}$ defines a probability density on the torus (or on $\Sigma$).
For any measurable subset $A\subset \t2$, we will denote its measure by 
$$
WH^{k,\ell}_{\psi}(A)=\int_A WH^{k,\ell}_{\psi}(x)\,dx\,.
$$
In the semiclassical limit, a sequence of 
coherent states $\set{|\bep'\cdot\bep\ra}$ can be associated with a single phase space point
$x\in\t2$ only if both sidelengths $D^{-\ell}$, $D^{k-\ell}$ of the associated rectangles 
decrease to zero. This is the case if and only if the index $\ell=\ell(k)$ is chosen to depend on $k$, in
the following manner:
\begin{equation}\label{e:semiclass}
\ell(k)\to\infty\quad \text{and}\quad k-\ell(k)\to\infty\quad\text{as}\quad k\to\infty\,.
\end{equation} 
Therefore, to define semiclassical
limit measures of sequences of eigenstates $\big(\psi_k\in \cH_{D^k}\big)_{k\to\infty}$, we will
consider sequences of Husimi representations $\big(WH^{k,\ell}\big)$ satisfying the
above conditions. For instance, 
we can consider the ``symmetric'' choice $\ell=\lfloor k/2 \rfloor$.

\subsubsection{Anti-Wick quantization of observables}\label{s:walsh-anti-wick}

In standard quantum mechanics, coherent states may also be used to quantize
observables (smooth functions on $\t2$), using the anti-Wick procedure.
In the Walsh framework, a similar (Walsh-)anti-Wick quantization can be defined,
but now it rather makes sense on observables $f$ on $\t2\simeq\Sigma$ 
which are Lipschitz-continuous with respect to the distance \eqref{e:metrics},
denoted by $f\in Lip(\Sigma)$.
The reason to choose this functional space (instead of some space of smooth functions
on $\t2$)
is that we want to prove Egorov's 
theorem, which involves both $f$ and its iterate $f\circ B$. It is therefore convenient
to require that both these functions belong to the same space
(we could also consider H\"older-continuous functions on $\Sigma$).

The Walsh-anti-Wick quantization is defined as follows.
For any $k$, one selects a family of quantum rectangles 
\eqref{e:q.rect.}, such that $\ell=\ell(k)$ satisfies the semiclassical 
condition \eqref{e:semiclass}. The
quantization of the observable $f$ is the following operator on $\cH_{D^k}$:
\begin{equation}\label{e:WaW-defi}
\Op_{k,\ell}(f)\defi D^k \sum_{[\bep'\cdot\bep]\in \cR^{k,\ell}} 
|\bep'\cdot\bep\ra\la\bep'\cdot\bep|\;\int_{[\bep'\cdot\bep]}f(x)\,dx
=\sum_{[\bep'\cdot\bep]\in \cR^{k,\ell}}|\bep'\cdot\bep\ra\la\bep'\cdot\bep|\;
\overline{f}^{[\bep'\cdot\bep]}\,.
\end{equation}
Here and in the following, 
we denote by $\overline{f}^R$ the average of $f$ over the rectangle $R$.
For each $\ell$, the above operators form a {\em commutative} algebra,
namely the algebra of diagonal matrices in the 
$\ell$-basis. 
The quantization $\Op_{k,\ell}$ is in some sense the dual of 
the Husimi representation
$WH^{k,\ell}$~:
\begin{equation}\label{e:Hus-antiW}
\forall f\in Lip(\Sigma),\ \forall \psi\in \cH_{D^k},\qquad
\la\psi|\Op_{k,\ell}(f)|\psi\ra = \int_{\t2}WH^{k,\ell}_\psi(x)\,f(x)\,dx\,.
\end{equation}

The following proposition shows that this family of quantizations satisfy 
a certain number of ``reasonable'' properties. We recall that the Lipschitz norm
of $f\in Lip(\Sigma)$ is defined as 
$$
\|f\|_{Lip}\defeq\sup_{x\in\Sigma}|f(x)|+\sup_{x\neq y\in\Sigma}\frac{|f(x)-f(y)|}{d_\Sigma(x,y)}\,.
$$
\begin{prop}\label{p:equiv}
i) For any index $0\leq \ell\leq k$ and observable $f\in Lip(\Sigma)$, one has
$$
\Op_{k,\ell}(f^*)=\Op_{k,\ell}(f)^*,\qquad
\tr\big(\Op_{k,\ell}(f)\big)=D^k\,\int_{\t2}f(x)\,dx\,.
$$

ii) For any $0\leq \ell\leq k$ and observables $f,g\in Lip(\Sigma)$, 
\begin{equation}\label{e:product}
\|\Op_{k,\ell}(f\,g)-\Op_{k,\ell}(f)\Op_{k,\ell}(g)\|\leq \|f\|_{Lip}\,\|g\|_{Lip}\,
D^{-\min(\ell,k-\ell)}\,.
\end{equation}

iii) For any pair of indices $0\leq \ell'\leq \ell\leq k$, the two quantizations
$\Op_{k,\ell}$, $\Op_{k,\ell'}$ are related as follows:
$$
\forall f\in Lip(\Sigma),\qquad \|\Op_{k,\ell}(f)-\Op_{k,\ell'}(f)\|
\leq 2\,\|f\|_{Lip}\,D^{-\min(\ell',k-\ell)}\,.
$$
\end{prop}
The first two statements make up the ``correspondence principle
for quantum observables'' of Marklof and O'Keefe \cite[Axiom 2.1]{MarOKee05},
which they use to prove Quantum Ergodicity (see Theorem~\ref{c:QE} below).

The third statement implies that if $\ell'\leq \ell$ (depending on $k$) both satisfy the 
semiclassical condition \eqref{e:semiclass}, then the two quantizations
are asymptotically equivalent. 
\begin{proof}
The statement $i)$ is obvious from the definition \eqref{e:WaW-defi} and the 
fact that $\ell$-coherent
states form an orthonormal basis.

To prove $ii)$ and $iii)$ we use the Lipschitz regularity of the observables.
The variations of $f\in Lip(\Sigma)$ inside a rectangle $R=[\bal'\cdot\bal]$ are
bounded as follows:
$$
\forall x,y\in R, \qquad |f(x)-f(y)|\leq \|f\|_{Lip}\, d_\Sigma(x,y)\leq 
\|f\|_{Lip}\, \diam(R)\,,
$$
where the diameter of the rectangle $R$ for the metrics $d_\Sigma$ is
$\diam(R)=D^{-\min(|\bal|,|\bal'|)}$.
As a consequence, 
\begin{equation}\label{e:Lip1}
\forall x\in R, \qquad\left| f(x)-\overline{f}^R\right|\leq
\|f\|_{Lip}\, \diam(R)\,.
\end{equation}
To show $ii)$, we expand the operator in the left hand side of \eqref{e:product}:
$$
\Op_{k,\ell}(f\,g)-\Op_{k,\ell}(f)\Op_{k,\ell}(g)=
\sum_{[\bep'\cdot\bep]\in \cR^{k,\ell}} 
|\bep'\cdot\bep\ra\la \bep'\cdot\bep|\;
\left(\overline{(fg)}^{[\bep'\cdot\bep]}-
\overline{f}^{[\bep'\cdot\bep]}\;\overline{g}^{[\bep'\cdot\bep]}\right)\,.
$$
Using \eqref{e:Lip1} for $R=[\bep'\cdot\bep]\in\cR^{k,\ell}$, 
we easily bound the terms on the right hand side:
$$
\forall [\bep'\cdot\bep]\in\cR^{k,\ell}\,,
\qquad \left|\overline{(fg)}^{[\bep'\cdot\bep]}-\overline{f}^{[\bep'\cdot\bep]}\;
\overline{g}^{[\bep'\cdot\bep]}\right| 
\leq \|f\|_{Lip}\,\|g\|_{Lip}\,D^{-\min(\ell,k-\ell)}\,.
$$
Since the $\ell$-coherent states are orthogonal, Pythagore's theorem
gives the bound \eqref{e:product}.

To prove the statement $iii)$, we need to consider
``mesoscopic rectangles'' of the type $R=[\bal'\cdot\bal]$ where $|\bal|=\ell'$, 
$|\bal'|=k-\ell$. 
Such a rectangle $R$ supports $D^{\ell-\ell'}$ quantum rectangles of type $\cR^{k,\ell}$, 
and the same number of rectangles of type $\cR^{k,\ell'}$. We want to analyze the
partial difference
\begin{equation}\label{e:diff}
\Delta\Op(f)_{|R}\defi 
\sum_{\begin{subarray}{l}
[\bep'\cdot\bep]\in \cR^{k,\ell}\\
[\bep'\cdot\bep]\subset R\end{subarray}} 
|\bep'\cdot\bep\ra\la\bep'\cdot\bep|\;\overline{f}^{[\bep'\cdot\bep]} 
-\sum_{\begin{subarray}{l}
[\bep'\cdot\bep]\in \cR^{k,\ell'}\\
[\bep'\cdot\bep]\subset R\end{subarray}} 
|\bep'\cdot\bep\ra\la\bep'\cdot\bep|\;\overline{f}^{[\bep'\cdot\bep]}\,.
\end{equation}
Both terms of the difference act inside the same subspace 
$$
V_R=\Span\set{|\bep'\cdot\bep\ra\,:\,[\bep'\cdot\bep]\in \cR^{k,\ell},
[\bep'\cdot\bep]\subset R }.
$$
We then use \eqref{e:Lip1}
to show that the average of $f$ over any quantum rectangle 
$[\bep'\cdot\bep]\subset R$ satisfies
$$
\left|\overline{f}^{[\bep'\cdot\bep]} - \overline{f}^R\right|\leq \|f\|_{Lip}\, \diam(R)\,.
$$
Inserted in \eqref{e:diff}, this estimate yields the upper bound:
$$
\|\Delta\Op(f)_{|R}\|\leq 2\,\|f\|_{Lip}\, \diam(R)\,.
$$
Finally, since the subspaces $V_R$, $V_R'$ 
associated with two disjoint rectangles $R\neq R'$ are
orthogonal, Pythagore's theorem implies the statement $iii)$.
\end{proof}

\subsection{Walsh-quantized baker}
We are now
in position to adapt the Balazs-Voros quantization of the $D$-baker's map \eqref{e:baker}
to the Walsh framework, by mimicking \eqref{e:Weyl-baker}. 
We define the Walsh quantization of $B$ by the following unitary matrix 
$B_k$ in the position basis:
\begin{equation}\label{eq:Walsh-baker}
B_k\defi W_{D^k}^{-1}
\begin{pmatrix}
W_{D^{k-1}} & 0 & 0\\
0 & \ddots & 0\\
0 & 0 & W_{D^{k-1}}
\end{pmatrix}\,.
\end{equation}
This operator acts simply on tensor product states:
\begin{equation}\label{eq:walsh-baker}
B_k(v^{(1)}\otimes\ldots\otimes v^{(k)})
=v^{(2)}\otimes v^{(3)}\otimes\ldots\otimes v^{(k)}\otimes\cF_{D}^{*}v^{(1)}\,.
\end{equation}
Similarly, a tensor-product operator on $\cH_{D^k}$ will be transformed as
follows by the quantum baker:
\begin{equation}\label{eq:action-observ}
B_{k}(A^{(1)}\otimes\ldots\otimes A^{(k)})B_{k}^{-1}
=A^{(2)}\otimes A^{(3)}\otimes\ldots\otimes A^{(k)}\otimes\cF_{D}^{*}A^{(1)}\cF_{D}\,.
\end{equation}
These formulas are clearly reminiscent of the shift \eqref{eq:baker-shift}
produced by the classical map. The main difference lies in the fact that ``quantum 
sequences'' are of finite length $k$, the shift acting cyclically on the sequence, 
and one needs to act with $\cF^{*}_D$ on the last qu$D$it. 

This quantization of
the baker's map has been introduced before, 
as the extreme member among a family of different quantizations
\cite{SchaCav00}, and some of its semiclassical 
properties have been studied in \cite{TracScot02}. In particular, it was
shown that, within the standard Wigner-Weyl formalism, this family of 
quantum propagators does not
quantize the baker's map, but a multivalued version of it.  

\medskip

On the other hand, in this paper we will stick to the Walsh-anti-Wick formalism to 
quantize observables, and in this setting
we prove in the next proposition that the 
quantum baker \eqref{eq:Walsh-baker} quantizes the original baker's map.
\begin{prop}[Egorov theorem]\label{p:Egorov}
Let us select a quantization $\Op_{k,\ell}$ satisfying the semiclassical 
conditions \eqref{e:semiclass}. Then, for any observable $f\in Lip(\Sigma)$, we have
in the semiclassical limit
$$
\|B_{k}^{-1} \Op_{k,\ell}(f) B_{k} - \Op_{k,\ell}(f\circ B)\|\leq 
2\,\|f\|_{Lip}\,D^{1-\min(\ell,k-\ell-1)}\,.
$$
For the ``symmetric'' choice $\ell=\lfloor k/2 \rfloor$, the right hand side is 
of order $D^{-k/2}\sim \hbar^{1/2}$.
\end{prop}
\begin{proof}
The crucial argument is the fact that, for any index $0<\ell\leq k$,
the Walsh-baker maps $\ell$-coherent states
onto $(\ell-1)$-coherent states. 
This fact is obvious from the
definition \eqref{e:CS} and the action of $B_{k}$ on tensor product states 
\eqref{eq:walsh-baker}:
\begin{equation}\label{e:CS-evol}
\forall [\bep'\cdot\bep]\in\cR^{k,\ell},\qquad B_{k}|\bep'\cdot\bep\ra = 
|B(\bep'\cdot\bep)\ra
=|\ep'_{k-\ell}\ldots\ep'_2\ep'_1\ep_1\,\cdot\,\ep_2\ldots\ep_\ell\ra\,.
\end{equation}
Notice that the shifted rectangle $B([\bep'\cdot\bep])\in \cR^{k,\ell-1}$.
As a result, the evolved operator $B_{k}^{-1} \Op_{k,\ell}(f) B_{k}$ 
will be a sum of terms of the form
$$
|B^{-1}(\bep'\cdot\bep)\ra\la B^{-1}(\bep'\cdot\bep)|\;\overline{f}^{[\bep'\cdot\bep]}
=|B^{-1}(\bep'\cdot\bep)\ra\la B^{-1}(\bep'\cdot\bep)|\;
\overline{f}^{B^{-1}([\bep'\cdot\bep])}\;,
$$
which implies the exact formula
\begin{equation}\label{e:obser-evol}
B_{k}^{-1} \Op_{k,\ell}(f) B_{k}=\Op_{k,\ell+1}(f\circ B)\,.
\end{equation}
The third statement of Proposition~\ref{p:equiv} and the inequality
$\|f\circ B\|_{Lip}\leq D\, \|f\|_{Lip}$ 
yield the estimate.
\end{proof}

\begin{rem} The exact evolution \eqref{e:CS-evol} is similar with the
evolution of Gaussian coherent states through quantum cat maps \cite{FNdB03}.
It is also the Walsh counterpart of the coherent state evolution through the
Weyl-quantized baker $B^{BV}_N$, used in \cite{DENW} to prove a weak version of Egorov's 
property. In that case, the coherent states needed to be situated ``far away'' 
from the discontinuities of
$B$, which implied that Egorov's property only held for observables vanishing in some 
neighbourhood of the discontinuities. 
In the present framework, we do not need to take care of discontinuities, since
$B$ is continuous in the topology of $\Sigma$.
\end{rem}

\begin{rem}
The integer $k$ satisfies $k=\frac{|\log h|}{\log D}$, where $h=h_k=D^{-k}$ is 
Planck's constant, and $\log D$ the uniform Liapounov exponent of the
classical baker's map: $k$ is the \emph{Ehrenfest time} 
for the quantum baker. 
As in the Weyl formalism \cite{DENW}, the 
Egorov property can be extended to iterates $(B_k)^n$ up to times 
$n\approx (1-\delta)\frac{k}{2}$, for any fixed $\delta>0$.
\end{rem}

The exact evolution of coherent states \eqref{e:CS-evol} also implies 
the following property, dual of Eq.~\eqref{e:obser-evol}:
$$
\forall\psi\in\cH_{D^k},\qquad WH^{k,\ell}_{B_{k}\psi}=WH^{k,\ell-1}_\psi\circ B^{-1}\,.
$$
In particular, if $\psi$ is an eigenstate of $B_{k}$, one has
$$
WH^{k,\ell}_{\psi}=WH^{k,\ell-1}_\psi\circ B^{-1}\,,
$$
meaning that the classical map sends one Husimi representation to the next one.

The Egorov estimate of Proposition~\ref{p:Egorov} leads to the following
\begin{cor}[Invariance of semiclassical measures]
Consider a semiclassical sequence $(\psi_{k}\in\cH_{D^k})_{k\in\IN_*}$ such that
each $\psi_{k}$ is an eigenstate of $B_{k}$. It induces a sequence of Husimi measures
$\big(WH^{k,\ell}_{\psi_k}\big)$, where $\ell=\ell(k)$ is assumed to satisfy \eqref{e:semiclass}.
Up to extracting a subsequence, one can assume that this sequence converges to
a probability measure $\mu$ on $\Sigma$.

Then the measure $\mu$ is invariant through the baker's map $B$.
\end{cor}
This measure $\mu$ projects to a measure on $\t2$,
which we will also (with a slight abuse) call $\mu$.
The proof of Quantum Ergodicity \cite{BouzDB96,Zel97}, 
starting from the ergodicity of the classical map
with respect to the Lebesgue measure, is also valid within our nonstandard
quantization. Indeed, as shown in \cite{MarOKee05}, 
the statements $i),ii)$ of Proposition~\ref{p:equiv} and
the Egorov theorem (Prop.~\ref{p:Egorov}) suffice to prove Quantum Ergodicity for 
the Walsh-quantized baker:
\begin{thm}[Quantum Ergodicity]\label{c:QE}
For any $k\in\IN_*$, select an orthonormal 
eigenbasis $(\psi_{k,j}\in\cH_{D^k})_{j=0,\ldots,D^k-1}$ of the 
Walsh-quantized baker $B_k$. 

Then, for any $k\geq 1$, there exists a subset $J_k\subset\set{0,\ldots,D^k-1}$ such that
\begin{itemize}
\item $\lim_{k\to\infty}\frac{\sharp J_k}{D^k}=1$ (``almost all eigenstates'')
\item if $\ell(k)$ satisfies \eqref{e:semiclass} and $j(k)\in J_k$ for all $k\geq 1$, 
then the sequence of 
Husimi measures $(WH^{k,\ell(k)}_{\psi_{k,j(k)}})$ weakly converges to the Lebesgue
measure on $\t2$.
\end{itemize}
\end{thm}

\begin{rem}\label{r:WL} 
In the following section we will be working with partitions
into the vertical rectangles $[\cdot\bal]$, $|\bal|=n$,
which make up the partition $\cP^{(n)}$ (see section~\ref{s:generating}). 
For any state $\psi\in\cH_{D^k}$, 
the measure $WH^{k,k}_\psi$ assigns the weight 
$|\la \bq_j|\psi\ra|^2$ to each vertical quantum rectangle 
$[\cdot\bep]$, $|\bep|=k$. With respect to the partition $\cP^{(n)}$,
all Husimi measures $WH^{k,\ell}_\psi$, $n\leq \ell \leq k$ are equivalent:
for any cylinder $[\cdot\bal]\in \cP^{(n)}$, we indeed have
\begin{equation}\label{e:inv2}
\forall \ell,\ n\leq\ell\leq k,\qquad 
WH^{k,\ell}_\psi([\cdot\bal])=WH^{k, k}_\psi([\cdot\bal])\,.
\end{equation}
\end{rem}

\section{Some explicit eigenstates of $B_{k}$} 
\label{sub:particular-eigenstates}
The interest of the quantization $B_{k}$ lies in the fact that its spectrum and
eigenstates can 
be analytically computed. 

\subsection{Short quantum period}\label{s:short-period}
The crucial point (derived from the identity 
\eqref{eq:walsh-baker} and the
periodicity of the Fourier transform) is that this operator is periodic,
with period $2k$ (when $D=2$) or $4k$ (when $D\geq 3$):
\begin{align*}
D=2&\Longrightarrow \forall k\geq 1,\quad (B_{k})^{2k}=I_{2^k}\\
D\geq 2&\Longrightarrow \forall k\geq 1,\quad (B_{k})^{4k}=I_{D^k}\,.
\end{align*}
More precisely, $(B_{k})^{2k}$ is the involution
\begin{equation}\label{e:parity}
(B_{k})^{2k}=\Pi\otimes\Pi\cdots\otimes \Pi\,,
\end{equation}
where $\Pi$ is the ``parity operator'' on $\IC^D$, which sends $e_{\ep}$ to
$e_{\bar{\ep}}$, with $\ep+\bar{\ep}\equiv 0\bmod D$.

As we noticed above,
$k=\frac{|\log h|}{\log D}$ is the Ehrenfest time of the system, so the above
periodicity can be compared with the ``short quantum periods'' of the quantum 
cat map \cite{BonDB2,FNdB03}, which allowed one to construct eigenstates with a 
partial localization on some periodic orbits.
The first consequence of this logarithmic period is the very high
degeneracy of the eigenvalues $\set{e^{2 i\pi r/4k},\ \ r=0,\ldots,4k-1}$:
each of them is approximately $\frac{D^k}{4k}$-degenerate. In the case of 
the cat map, this huge degeneracy gives sufficient freedom to
construct eigenstates which are partially scarred on
a periodic orbit \cite{FNdB03}. 
In the Walsh-baker case, 
although $4k$ is the {\em double} of what 
was called a ``short period'' in \cite{FNdB03}, $(B_{k})^{2k}$
sends a coherent state $|\bep'\cdot\bep\ra$ to another coherent state 
$|\bar{\bep}'\cdot\bar{\bep}\ra$, and we are still able to construct
half-scarred eigenstates. Due to \eqref{e:parity}, 
a state scarred on 
the periodic orbit indexed by the periodic sequence
$(\ep_1\ep_2\ldots\ep_p)$ is also scarred, with the same
weight, on the ``mirror'' orbit $(\bar\ep_1\bar\ep_2\ldots\bar\ep_p)$.

\begin{figure}
\includegraphics[angle=-90,width=7.3cm]{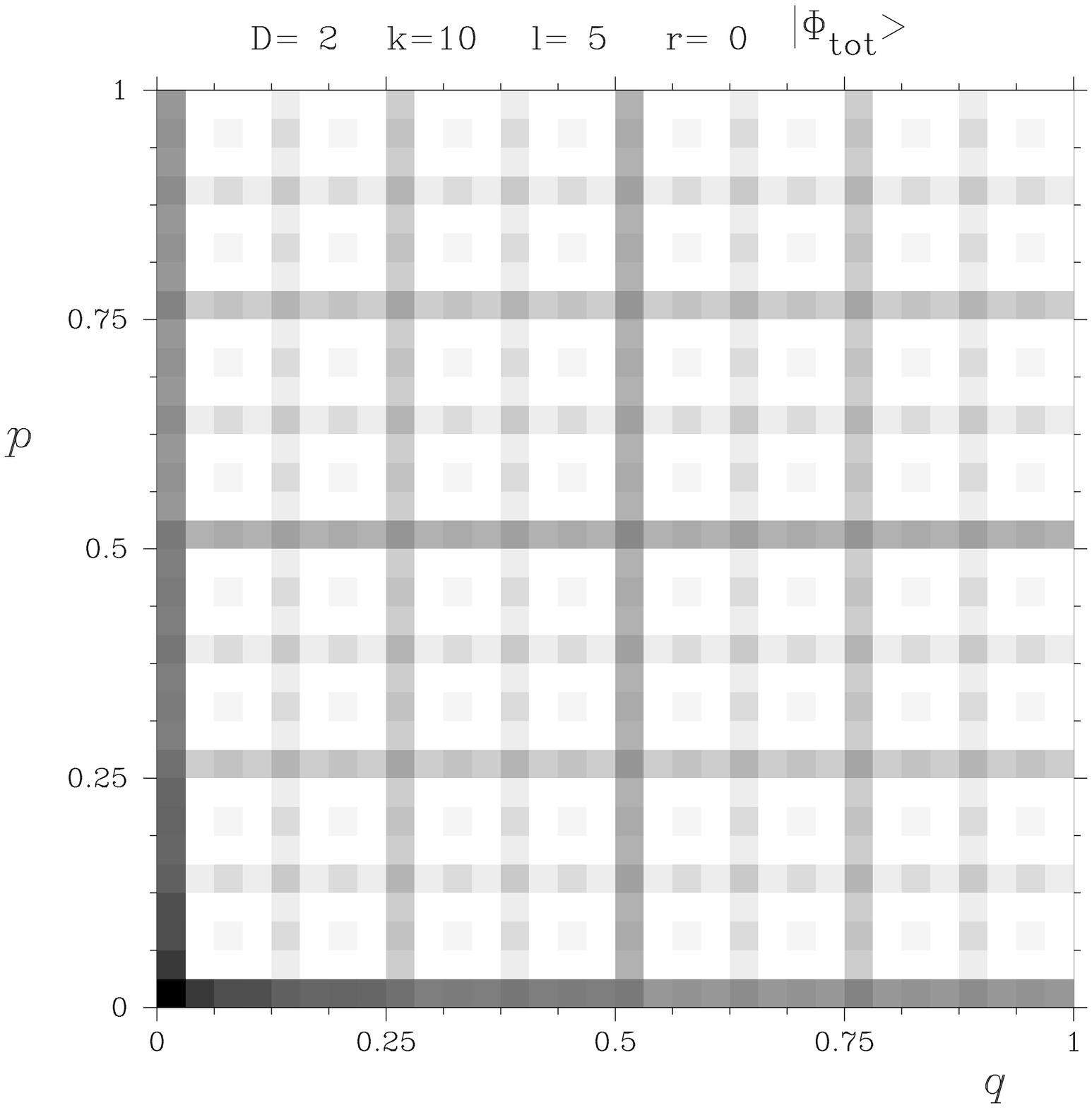}
\includegraphics[angle=-90,width=7.3cm]{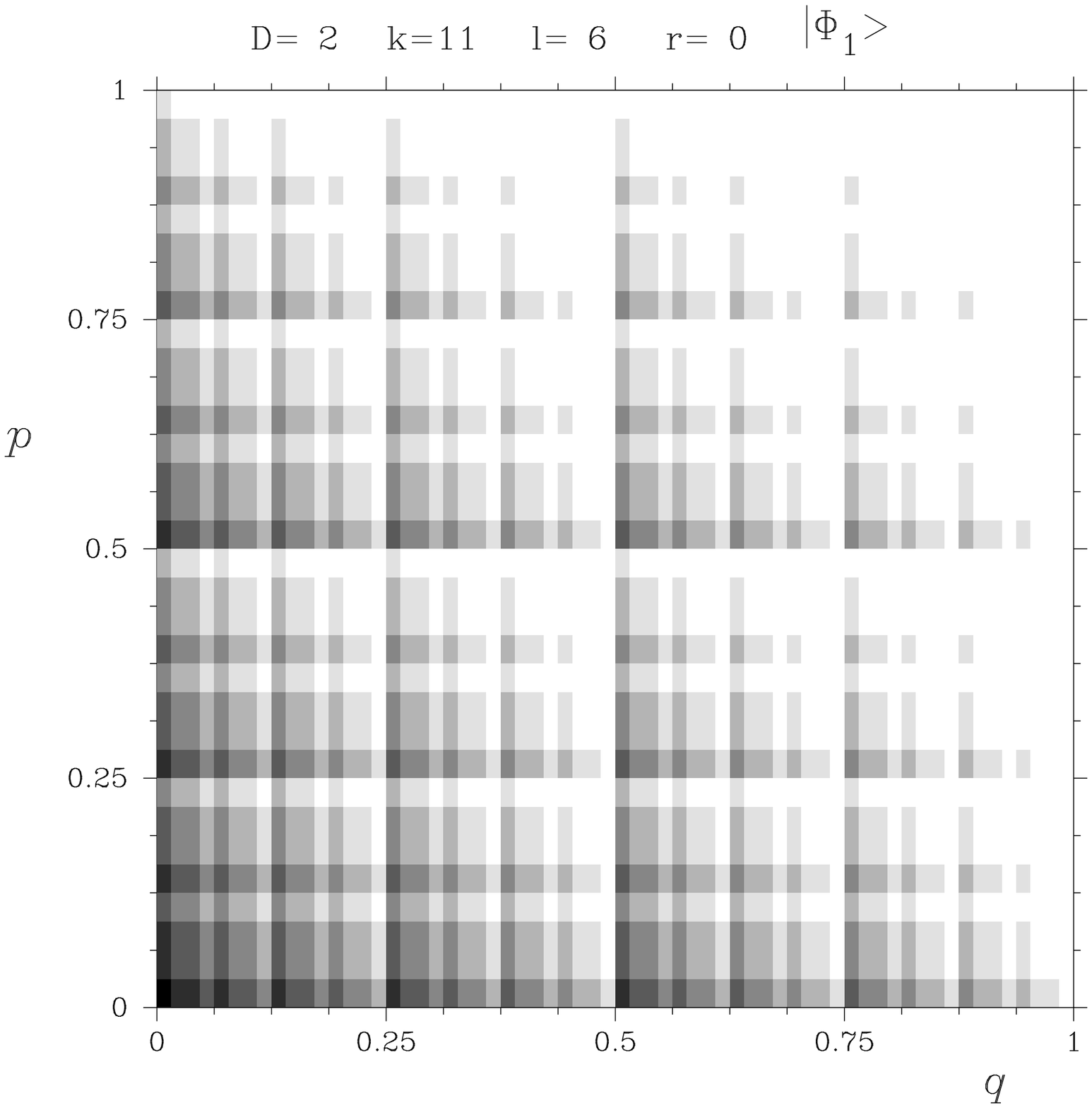}\\
\includegraphics[angle=-90,width=7.3cm]{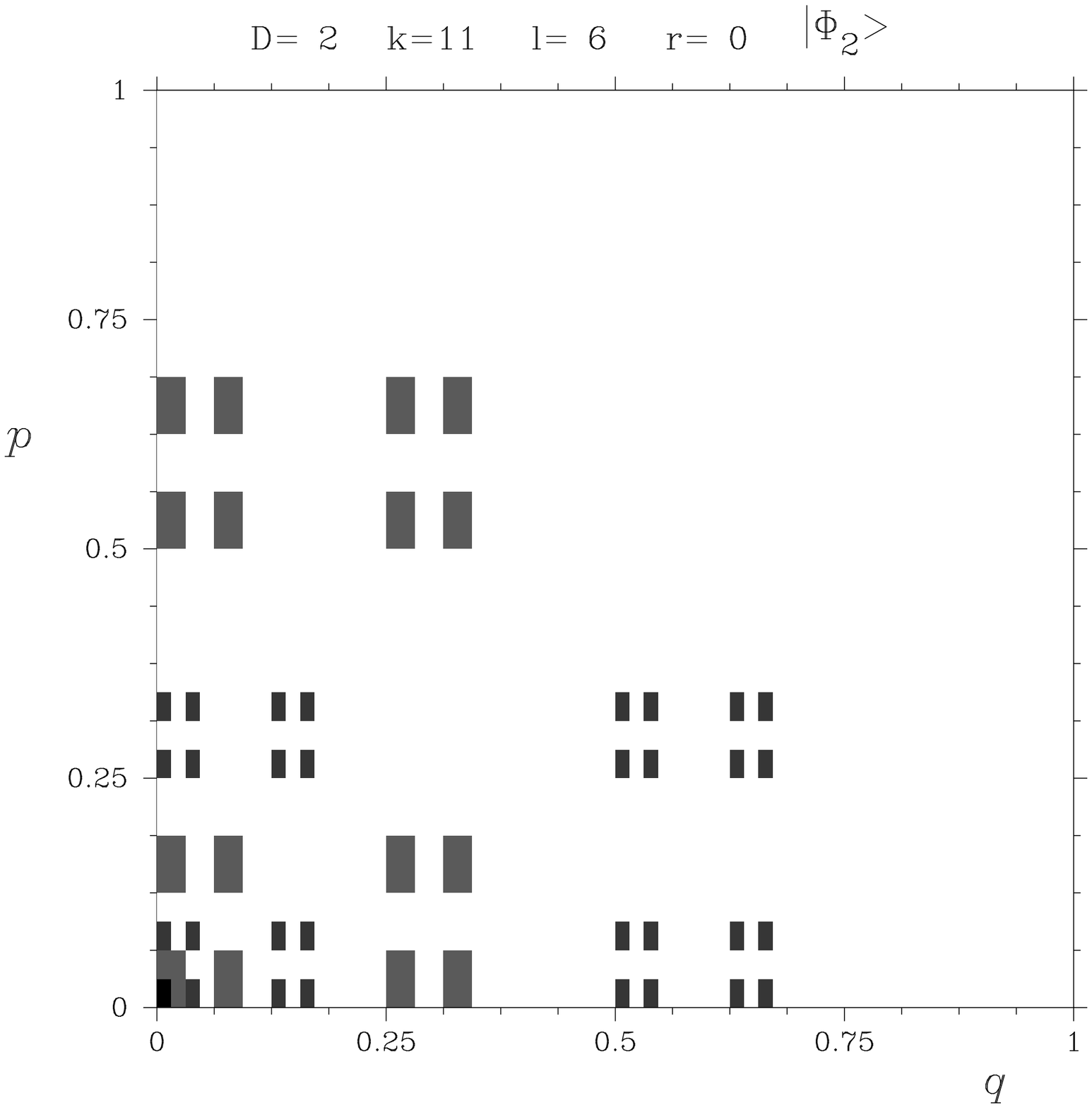}
\caption{Eigenstates of $B_k$ for $D=2$. The grey
scale corresponds to a logarithmic representation of $WH^{k,\ell}_{\psi}(x)$ 
(black=large; white=small). Top left: eigenstate half-scarred at the origin, $k=10$, $\ell=5$. 
Top right: tensor product eigenstate \eqref{eq:tensor-eigenstate}, $k=11$, $\ell=6$. 
Bottom: eigenstate \eqref{e:example3} with a fractal support, $k=11$, $\ell=6$ (white=zero).}
\label{fig:husimis}
\end{figure}

\subsection{Tensor-product eigenstates}\label{s:tensor-prod}
A new feature, compared with the quantum cat map, is that we
straightforwardly obtain eigenstates of $B_k$ which are not ``scarred'' on any
periodic orbit, but still have
a nontrivial phase space distribution: 
the associated semiclassical measure is a singular Bernoulli measure.
These states are constructed as follows: 
take any eigenstate $w\in \IC^D$ of the inverse Fourier transform
$\cF_D^*$. 
Then, for any $k\geq 1$, 
the tensor-product state 
\begin{equation}\label{eq:tensor-eigenstate}
\psi=w\otimes\cdots\otimes w\in \cH_{D^k}
\end{equation}
is an eigenstate of $B_{k}$. From \eqref{eq:Walsh-Husimi}, its Husimi measure 
$WH^{k,\ell}_{\psi}$ has the following weight on a quantum 
rectangle $[\bep'\cdot\bep]\in\cR^{k,\ell}$:
\begin{equation}\label{eq:Husimi-fractal}
WH^{k,\ell}_{\psi}([\bep'\cdot\bep])=
|w_{\ep_1}|^2\ldots |w_{\ep_{\ell}}|^2\: |w_{\ep'_1}|^2\ldots 
|w_{\ep'_{k-\ell}}|^2\,.
\end{equation}
This shows that $WH^{k,\ell}_{\psi}$ is the product of a measure $\nu_\ell$ on the horizontal
interval by a measure $\nu_{k-\ell}$ on the vertical interval. $\nu_\ell$ (resp.
$\nu_{k-\ell}$) can be obtained by conditioning a certain self-similar measure $\nu$ 
on subintervals of type $[\frac{j}{D^\ell},\frac{j+1}{D^\ell})$
(resp. $[\frac{j}{D^{k-\ell}},\frac{j+1}{D^{k-\ell}})$).
This measure $\nu$ is constructed by iteration: the first step
consists in splitting $[0,1)$ into $D$ subintervals $[\frac{\ep}{D},\frac{\ep+1}{D})$, 
and allocating the weight
$p_{\ep}=|w_\ep|^2$ to the $\ep$-th subinterval. The next step splits each subinterval, etc.
In other words, for any finite sequence $\bep\in\Sigma_n$, 
the measure of the interval $[\cdot\bep]$ is given by
$$
\nu([\cdot\bep])=p_{\ep_1}p_{\ep_2}\ldots p_{\ep_n}\,.
$$
In the symbolic representation $[0, 1)\sim \Sigma_+$,
$\nu$ is a Bernoulli measure.

The Husimi measure $WH^{k,\ell}_{\psi}$ is therefore the measure
$$
\mu=\nu(dq)\times \nu(dp),\quad\text{conditioned on the rectangles }
[\bep'\cdot\bep]\in\cR^{k,\ell}\,.
$$
Assuming that $\ell$ satisfies the condition \eqref{e:semiclass} (so that
the diameters of the rectangles vanish as $k\to\infty$),
we get
$$
\lim_{k\to\infty}WH^{k,\ell}_{\psi}=\mu\,,
$$
where the limit should be understood in the weak sense.

The measure $\mu$ is obviously a Bernoulli invariant measure, of the type shown in
the Examples of Section~\ref{s:generating}.
Let us describe some particular cases, forgetting for a moment that the state $w$ is
an eigenstate of $\cF^*_D$, and taking for $w$ any normalized state in $\IC^D$.
\begin{itemize}
\item if the coefficients $p_\ep$ are all equal, $p_\ep=1/D$,
then $\mu=dx$ is the Lebesgue measure.  
\item if there is a single $\ep\in\set{0,\ldots,D-1}$ such that $p_\ep=1$ 
and the others vanish,
then $\mu=\delta_{x_o}$, where $x_o\equiv \ldots\ep\ep\,\cdot\,\ep\ep\ep\ldots$
is a fixed point of $B$.
Obviously, this is impossible if $w$ an eigenstate of $\cF^*_D$.
\item in the remaining cases, $\mu$ is a purely singular continuous 
measure on $\t2$, with simple self-similarity  properties.
\end{itemize}

\subsubsection*{Topological entropy of tensor product eigenstates}
An eigenstate $w$ of $\cF^*_D$ can have
a certain number of vanishing coefficients. Call 
$S\subset \set{0,\ldots,D-1}$ the set of non-vanishing coefficients, and
$d=\sharp S$ its cardinal. If $d<D$, the corresponding
measure $\mu$ is then supported on a proper invariant
subset $F_\mu$ of $\t2$, corresponding
to the sequences $\bep'\cdot\bep\in\Sigma$ with all 
coefficients $\ep_i,\ep'_i\in S$. 
One can easily check that the topological entropy of $F_\mu$ is given by
$$
h_{top}(F_\mu)=\log (d)\,.
$$
Now, because all the 
matrix elements of $\cF^*_D$ are of modulus $D^{-1/2}$, the number
$d$
of non-vanishing components of $w$ is bounded as
\begin{equation}\label{e:htop-bound}
d\geq\sqrt{D}\,,\qquad\mbox{so that}\quad h_{top}(F_\mu)\geq \frac{\log D}{2}\,.
\end{equation}
This proves that semiclassical measures $\mu$ obtained from sequences of
tensor-product eigenstates \eqref{eq:tensor-eigenstate} 
satisfy the general lower bound of Theorem~\ref{HTOP}.

The simplest example of such eigenstates seems to be for $D=4$: 
$\cF_{4}^{*}$
admits the eigenstate $w=(1,0,1,0)/\sqrt{2}$.
The corresponding limit measure $\mu$ is supported on a subset $F_\mu$ which
saturates the lower bound \eqref{e:htop-bound}: $h_{top}(F_\mu)=\log 2=\frac{\log 4}{2}$.

\subsubsection*{Metric entropy of tensor product eigenstates}
For a normalized state $w\in\IC^D$, the Kolmogorov-Sinai entropy of the 
measure $\mu$ can be shown to be
$$
h_{KS}(\mu)=-\sum_{\ep=0}^{D-1} p_{\ep}\log p_{\ep}
=-\sum_{\ep=0}^{D-1}|w_{\ep}|^{2}\log|w_{\ep}|^{2}\defeq h(w).
$$
A priori, this function could take any value between $0$ and $\log D$, the topological
entropy of $\t2$ with respect to the baker's map. However, as in the case of the
topological entropy, imposing $w$ to be an eigenstate of $\cF^*_D$ restricts
the possible range of $h(w)$. Indeed, the following
``Entropic Uncertainty Principle'', first conjectured by Kraus \cite{Kraus87} and
proven in \cite{MaaUff}, direcly provides the desired lower bound for $h(w)$.
\begin{thm}[Entropic Uncertainty Principle \cite{MaaUff}]
\label{thm:entropic_uncertainty}
For any $M\in\IN_*$, let $U$ be a unitary $M\times M$
matrix and $c(U)\defi\sup_{i,j}|U_{ij}|$. Then, for any
normalized state $\psi\in\IC^{M}$, one has
$$
h(\psi)+h(U\psi)\geq - 2\log c(U)\,,
$$
where the entropy is defined as $h(\psi)=-\sum_i |\psi_i|^2\log|\psi_i|^2$.
\end{thm}
The proof of this theorem 
(which is the major ingredient in the proof of Theorem \ref{HMET},
see Section~\ref{s:HMET}) is outlined in the Appendix.

Applying this theorem to the matrix $U=\cF_{D}^{*}$, and using the fact that $w$ is an
eigenstate of that matrix, we obtain the desired lower bound
\begin{equation}\label{e:HMET-tensor}
h_{KS}(\mu)=h(w)\geq\frac{\log D}{2}.
\end{equation}
The above example of tensor-product eigenstates of the $4$-baker, 
constructed from $w=(1,0,1,0)/\sqrt{2}$, also saturate
this inequality:
$h_{KS}(\mu)=h_{top}(F_{\mu})=\frac{\log 4}{2}$. 

\subsection{A slightly more complicated example}
In the case of $D=2$, although none of the eigenvectors of $F_2$ has
any vanishing component, one can still construct eigenstates
converging to a fractal measure supported on a proper subset of $\t2$. Indeed, we
notice that $\cF_{2}\,e_{0}=\frac{e_{0}+e_{1}}{\sqrt{2}}\defi e_{+}$,
and $\cF_{2}^{2}=I_{2}$. As a result, in the case $k$ is odd, the
state 
\begin{equation}\label{e:example3}
\psi=\frac{1}{\sqrt{2}}\left(e_{0}\otimes e_{+}\otimes e_{0}\otimes\ldots e_{+}\otimes e_{0} + 
e_{+}\otimes e_{0}\otimes e_{+}\otimes\ldots e_{0}\otimes e_{+}\right)
\end{equation}
is an eigenstate of $B_{k}$. It becomes normalized in the limit $k\to\infty$,
and one can check that the associated semiclassical measure 
is
$\mu=\frac{1}{2}\left(\nu_{1}(dq)\times\nu_{2}(dp)+\nu_{2}(dq)\times\nu_{1}(dp)\right)$,
where $\nu_{1}$ (resp. $\nu_{2}$) is the self-similar measures on $[0,1)$
obtained by splitting $[0,1)$ in $4$ equal subintervals, which are allocated the
weights $(1/2,1/2,0,0)$ (resp. $(1/2,0,1/2,0)$), and so on.
One can easily show that this semiclassical
measure $\mu$ saturates both lower bounds: 
$h_{KS}(\mu)=h_{top}(F_\mu)=\frac{\log2}{2}$.

\section{Proof of Theorem \ref{HMET}: Lower bound on the metric entropy}\label{s:HMET}

Applying Theorem \ref{thm:entropic_uncertainty} in a more clever way,
we can generalize the lower bound \eqref{e:HMET-tensor} to any 
semiclassical measure $\mu$, thereby proving Theorem \ref{HMET}. In this section we give
ourselves a sequence $\big(\psi_k\in\cH_{D^k}\big)$ of eigenstates of $B_k$, and assume that 
the associated Husimi measures
converge to an invariant probability measure $\mu$.

\subsection{Quantum partition of unity}
The definition of metric entropy given in Section~\ref{s:metric} starts from 
the ``coarse'' partition $\cP$ (made of $D$ rectangles $[\cdot\ep]$), 
which is then refined into a sequence of partitions
$\cP^{(n)}$ using the classical dynamics. A natural way to study the Kolmogorov-Sinai entropy of
quantum eigenstates is to transpose these objects to the quantum framework. For any anti-Wick
quantization $\Op_{k,\ell}$ satisfying the condition \eqref{e:semiclass}, 
the characteristic functions $\bbbone_{[\cdot \ep]}$ are quantized into the orthogonal 
projectors
\begin{equation}\label{e:P_eps}
P_{\ep}=\pi_{\ep}\otimes (I)^{\otimes k-1}\,,\qquad\ep=0,\ldots,D-1\,.
\end{equation}
Here, $\pi_\ep$ is the orthogonal projector on the basis state $e_\ep\in\IC^D$, and $I=I_D$ is
the identity operator on $\IC^D$. 
This family of projectors make up a ``quantum partition of unity'':
$$
\sum_{\ep=0}^{D-1}P_\ep = (I)^{\otimes k}=I_{D^k}\,.
$$
Like its classical counterpart, this partition can be refined using the dynamics. To 
an evolved rectangle $B^{-l}([\cdot\ep])$ corresponds the projector
$$
P_{\ep}(l)\defi B_k^{-l}\,P_{\ep}\,B_k^{l}\,.
$$
From there, the quantum counterpart of the refined
partition $\cP^{(n)}=\set{[\cdot\bep],\:\, \bep\in\Sigma_n}$ 
is composed of the following operators:
\begin{equation}\label{e:B_bep}
P_{\bep}\defi P_{\ep_n}(n-1)\circ\ldots\circ P_{\ep_2}(1)\circ P_{\ep_1}\,.
\end{equation}
Using the formula \eqref{eq:action-observ}, we find that
\begin{equation}\label{e:n<k}
n\leq k\Longrightarrow P_{\bep}=
\pi_{\ep_1}\otimes\pi_{\ep_2}\otimes\ldots\pi_{\ep_n}\otimes (I)^{\otimes k-n}\,.
\end{equation}
This shows that $P_{\bep}$ is an orthogonal projector associated with the
rectangle $[\cdot\bep]$. It is equal to $\Op_{k,\ell}(\bbbone_{[\cdot\bep]})$ 
if $n\leq\ell$.
In the extreme case $n=k$, these operators project on single position eigenstates:
$$
\forall j=\ep_1\ldots\ep_k\in \set{0,\ldots,D^k-1},\qquad
P_{\bep}=|\bq_j\ra\la \bq_j|\,.
$$
Using Remark~\ref{r:WL}, we see that 
these projectors can be direcly used to express the weight of the 
Husimi measures on rectangles. Indeed, if $n\leq\ell\leq k$ and $[\cdot\bep]\in \Sigma_n$, then 
\begin{equation}\label{e:sq-norm}
WH^{k,\ell}_{\psi_k}([\cdot\bep])=\Vert P_{\bep}\, \psi_k\Vert^2\,.
\end{equation}
From there, we straightforwardly deduce the:
\begin{lem}\label{l:entropy}
Provided $n\leq\ell\leq k$, the entropy \eqref{e:def-entropy}
of the Husimi measure $WH^{k,\ell}_{\psi_k}$, 
relative to the refined 
partition $\cP^{(n)}$, can be written as follows:
\begin{equation}\label{e:n-k-l}
h_{\cP^{(n)}}(WH^{k,\ell}_{\psi_k})= -\sum_{|\bep|=n} 
\|P_{\bep}\,\psi_k\|^2\,\log \big(\|P_{\bep}\,\psi_k\|^2\big)\,.
\end{equation}
\end{lem}

For some values of the indices, this quantity corresponds to well-known 
``quantum entropies''.

\subsection{Shannon and Wehrl entropies}\label{s:Wehrl}

By setting $n=\ell=k$ in the above Lemma, we obtain a ``quantum'' entropy
which has been used before to characterize the localization properties of {\em individual} 
states \cite{Izra90}. 
It is simply the {\em Shannon entropy} of the state $\psi\in\hn$, when expressed 
in the position basis $\set{ \bq_{j},\; j=0,\ldots,N-1}$:
\begin{equation}\label{e:Shannon}
h_{Shannon}(\psi)\defeq h_{\cP^{(k)}}(WH^{k,k}_{\psi})=
-\sum_{j=0}^{N-1}|\la \bq_j|\psi\ra|^2\,\log\big(|\la \bq_j|\psi\ra|^2\big)\,.
\end{equation}
This entropy obviously selects a preferred ``direction'' in phase space: one could
as well consider the Shannon entropy in the momentum basis. To avoid this type of
choice, it has become more fashionable to use a quantum entropy based on the Husimi
representation of quantum states, introduced by Wehrl \cite{Wehrl}. In the
Weyl framework, it is given by the integral over the phase space of $\eta(|\la x|\psi\ra|^2)$,
where $\eta(s)=-s\log s$, and $\set{|x\ra\,:\,x\in\t2}$ is a continuous family of
Gaussian coherent states.

In the Walsh framework, the coherent states form  discrete families, so the integral
is effectively a sum. For any index $\ell$, we define the Walsh-Wehrl entropy of 
$\psi\in\cH_{D^k}$ as:
\begin{equation}\label{e:Wehrl}
h^{k,\ell}_{Wehrl}(\psi)=-\sum_{[\bep'\cdot\bep]\in \cR^{k,\ell}}
|\la \bep'\cdot\bep|\psi\ra|^2\,\log\big(|\la\bep'\cdot\bep|\psi\ra|^2\big)\,.
\end{equation}
Notice that
the Shannon entropy \eqref{e:Shannon} is a particular case of 
the Wehrl entropy, obtained by setting $\ell=k$.
Eq.~\eqref{e:CS-evol} implies that
all quantum entropies of eigenstates are equal:
\begin{prop}\label{p:Shannon=Wehrl}
If $\psi_k\in\cH_{D^k}$ is an eigenstate of the Walsh-baker $B_k$, then 
its Wehrl and Shannon entropies are all equal:
$$
\forall \ell\in[0,k], \qquad h^{k,\ell}_{Wehrl}(\psi_k)=h_{Shannon}(\psi_k)\,.
$$
\end{prop}

As in the case of Gaussian coherent states \cite{Wehrl,Lieb78}, 
localized states have a small Wehrl
entropy: the minimum of $h^{k,\ell}_{Wehrl}(\psi)$ is reached for
$\psi= |\bep'\cdot\bep\ra$ a coherent state in the $\ell$-basis,
where the entropy vanishes.
On the opposite, the entropy is maximal when $\psi$ is equidistributed
with respect to the $\ell$-basis,
and the entropy then takes the value $\log N=|\log 2\pi\hbar|$. 
Notice that the extremal properties
of the entropy $h^{k,\ell}_{Wehrl}$ of pure quantum states 
are much easier to analyze than those of the ``Gaussian''
Wehrl entropies on the plane, the torus or the sphere \cite{Lieb78,Lee88,NV98}.

The Shannon or Wehrl entropies 
can be now bounded from below using the Entropic Uncertainty Principle,
Theorem~\ref{thm:entropic_uncertainty}.
Indeed, $\psi_k$ is an eigenstate of the iterate 
$(B_k)^k$, which is the tensor product operator
\begin{equation}\label{e:power-k}
(B_k)^k=\cF_{D}^{*}\otimes\cF_{D}^{*}\otimes\ldots\otimes\cF_{D}^{*}\,.
\end{equation}
The matrix elements of this operator in the position basis are all
of modulus $D^{-k/2}$. Thus, Theorem~\ref{thm:entropic_uncertainty}
implies that
\begin{equation}\label{e:EIP}
h_{Shannon}(\psi_k)=h_{\cP^{(k)}}(WH^{k,k}_{\psi_k}) \geq \frac{k}{2}\log D\,.
\end{equation}
Using the property that the Wehrl entropies \eqref{e:Wehrl} of an eigenstate are all equal
to each other (see Proposition \ref{p:Shannon=Wehrl}), this 
proves Theorem~\ref{t:Wehrl}.

In the expression for the Shannon entropy, both
the Husimi measure $WH^{k, \ell}_{\psi_k}$ and the partition $\cP^{(n)}$ depend on
the semiclassical parameter $k$ in a rigid way, namely $\ell=n=k$.
On the other hand, if we want to understand the entropy of the semiclassical measure $\mu$,
we should first estimate the entropy of some $k$-independent partition $\cP^{(n)}$,
then take the semiclassical limit ($k\to \infty$) of the Husimi measures 
$WH^{k,\ell}_{\psi_k}$ with the condition \eqref{e:semiclass} satisfied, and
only send $n$ to infinity afterwards.
In other words, we need to control the entropies \eqref{e:n-k-l} for a fixed
$n\in\IN$ while sending $k,\ell\to\infty$.

In the following sections, we present two different approaches to 
realize this program, both yielding a proof of Theorem~\ref{HMET}.

\subsection{First method: use of subadditivity}
The first approach consists in estimating the entropy \eqref{e:n-k-l} of the 
partition $\cP^{(n)}$ for some fixed $n$, starting from the lower bound \eqref{e:EIP} on 
the entropy 
of $\cP^{(k)}$. Both these entropies are taken on the measure $\mu_k\defeq WH^{k,k}_{\psi_k}$.
This estimation uses the subadditivity property \eqref{e:subadd}. 

Using Euclidean division, we can write $k=qn+r$ with $q, r\in\IN$, $r<n$.
The subadditivity of entropy implies that
\begin{equation}\label{e:subadd1}
h_{\cP^{(k)}}(\mu_k)\leq h_{\cP^{(n)}}(\mu_k)+h_{B^{-n}\cP^{(n)}}(\mu_k)
+\ldots+h_{B^{-(q-1)n}\cP^{(n)}}(\mu_k)+ h_{B^{-qn}\cP^{(r)}}(\mu_k).
\end{equation}
The very last term, being the entropy of a partition of $D^r$ elements, is less than
$r\log D$.

Using the fact that $\psi_k$ is an eigenstate of $B_{k}$, we prove below 
that the Husimi measure $\mu_k$ is invariant under $B$ until the Ehrenfest time~:
\begin{lem}\label{lem:invar2}
For any $n$-rectangle $[\cdot\bep]$ of the partition $\cP^{(n)}$,
for any index $0\leq l\leq k-n$,
we have
$$
\mu_k(B^{-l}[\cdot\bep])=\mu_k([\cdot\bep])\,.
$$
\end{lem}
This straighforwardly implies the following property:
$$
l\leq k-n\Longrightarrow h_{B^{-l}\cP^{(n)}}(\mu_k)=h_{\cP^{(n)}}(\mu_k)\,.
$$
Injecting this equality in the subadditivity \eqref{e:subadd1}, and 
using the lower bound \eqref{e:EIP} for $h_{\cP^{(k)}}(\mu_k)$, 
we obtain a lower bound for the entropy of the fixed partition $\cP^{(n)}$:
\begin{equation}\label{e:lower}
h_{\cP^{(n)}}(\mu_k)\geq \frac1{q}\Big(h_{\cP^{(k)}}(\mu_k)- r\log D\Big)\geq 
\frac1{q}\Big(k\frac{\log D}{2}-r\log D\Big)\,.
\end{equation}
From the identity \eqref{e:inv2}, and assuming that $\lfloor k/2 \rfloor>n$,
the left hand side is also 
the entropy of the Husimi measure $WH^{k,\lfloor k/2 \rfloor}_{\psi_k}$, 
which converges to $\mu$ in
the semiclassical limit. On the right hand side,  $k/q\to n$ and $r/q\to 0$ as
$k\to\infty$, so in the limit,
$$
h_{\cP^{(n)}}(\mu)
\geq\frac{n}{2}\,\log D\,.
$$
We can finally let $n\to \infty$, and get Theorem~\ref{HMET}.\qed

\subsubsection*{Proof of Lemma~\ref{lem:invar2}}
For any $n$-rectangle $[\cdot\bep]$ of the partition $\cP^{(n)}$,
we have
\begin{align*}
\mu_k([\cdot\bep])&=
\left\Vert \pi_{\ep_1}\otimes\ldots\pi_{\ep_n}\otimes (I)^{\otimes k-n}\psi_k\right\Vert^2\\
&=
\left\Vert (B_k)^{-l}\,
\big(\pi_{\ep_1}\otimes\ldots\pi_{\ep_n}\otimes (I)^{\otimes k-n}\big)\,(B_k)^l
\psi_k\right\Vert^2,
\end{align*}
where we have used the facts that $\psi_k$ is an eigenfunction of $B_k$, and
that $B_k$ is unitary. Now, using \eqref{eq:action-observ}, the last line can be transformed into
\begin{align*}
\left\Vert (I)^{\otimes l}\otimes\pi_{\ep_1}\otimes\ldots\pi_{\ep_n}
\otimes (I)^{\otimes k-n-l}\,\psi_k\right\Vert^2
&=\sum_{\alpha_1,\ldots,\alpha_l=0}^{D-1} 
\left\Vert \pi_{\alpha_1}\otimes\ldots \pi_{\alpha_l}\otimes\pi_{\ep_1}\otimes\ldots\pi_{\ep_n}
\otimes (I)^{\otimes k-n-l}\,\psi_k\right\Vert^2\\
=\sum_{\alpha_1,\ldots,\alpha_l} \left\Vert P_{\bal\bep}\psi_k\right\Vert^2
&=\sum_{\bal=(\alpha_1,\ldots,\alpha_l)} \mu_k([\cdot\bal\bep])=\mu_k(B^{-l}[\cdot\bep])\,.
\end{align*}
The last equality is due to the fact that the set $B^{-l}[\cdot\bep]$ is the disjoint union
\begin{equation}\label{eq:back}
B^{-l}[\cdot\bep]=\bigcup_{\alpha_1,\ldots,\alpha_l=0}^{D-1} [\cdot\bal\bep]\,.
\end{equation}\qed

\subsection{Second method: vectorial Entropic Uncertainty Principle}
\label{ss:GEUP}

The second approach to bound \eqref{e:n-k-l} from below is to directly apply to that sum the
vectorial version of the
Entropic Uncertainty Principle, given in Theorem~\ref{a:EUP} in the Appendix.

Indeed, for any $n\leq k$, the family of orthogonal projectors
$\set{P_{\bep},\;|\bep|=n}$ satisfy $P_{\bep}P_{\bep'}=\delta_{\bep\bep'}P_{\bep}$,
and the resolution of unity
$$
\sum_{|\bep|=n}P_{\bep}=I\,.
$$ 
Any state $\psi\in {\cH}_{D^k}$ can be decomposed into the sequence of
states $\set{\psi_{\bep}=P_{\bep}\psi,\;|\bep|=n}$, in terms of which
the entropy \eqref{e:n-k-l} can then be written as
\begin{equation}\label{e:entropy-Psi}
h_{\cP^{(n)}}(\mu_k)= -\sum_{|\bep|=n}\Vert\psi_{\bep} \Vert^2\;\log\Vert\psi_{\bep}\Vert^2\defi
h_n(\psi)\,.
\end{equation}
The vectorial Entropic Uncertainty Principle 
(Theorem \ref{a:EUP}), specialized to this family of orthogonal projectors,
reads as follows~:
\begin{thm}
\label{thm:entropic_uncertainty_tensored}
For a given $n\leq k$, and
any normalized state $\psi\in {\cH}_{D^k}$, let us define the entropy
$$
h_n(\psi)=-\sum_{|\bep|=n}\Vert\psi_{\bep} \Vert^2\;\log\Vert\psi_{\bep}\Vert^2\,. 
$$
Let $U$ be a unitary operator on $\cH_{D^k}$. 
For any sequences $\bep$, $\bep'$ of length $n$,
we call $U_{\bep, \bep'}=P_{\bep} U P_{\bep'}$, and 
$c_n(U)=\sup_{|\bep|=| \bep'|=n}\|U_{\bep, \bep'}\|$.

Then, for any normalized state $\psi\in {\cH}_{D^k}$,
one has
$$
h_n(\psi)+h_n(U\psi)\geq-2\log c_n(U).
$$
\end{thm}
We apply this theorem to the eigenstates 
$\psi_k\in {\cH}_{D^k}$,
using the operator $U=(B_k)^k$. It gives a lower bound 
for the entropy of the Husimi measure $\mu_k$~:
$$
h_{\cP^{(n)}}(\mu_k)\geq -\log c_n(U).
$$
To compute $c_n(U)$, we expand the operators $U_{\bep, \bep'}$ as tensor products, 
using (\ref{e:n<k},\ref{e:power-k}):
$$
U_{\bep, \bep'}=P_{\bep}(B_k)^k P_{\bep'}=\pi_{\ep_1}\cF^*\pi_{\ep_1'}\otimes
\pi_{\ep_2}\cF^*\pi_{\ep_2'}\otimes\ldots\otimes \pi_{\ep_n}\cF^*\pi_{\ep_n'}\otimes
\cF^*\otimes\ldots\otimes\cF^*\,.
$$
Each of the first $n$ tensor factors can be written as
$$
\pi_{\ep_i}\cF^*\pi_{\ep_i'}=\cF^*_{\ep_i\ep'_i}\,|e_{\ep_i}\ra\la e_{\ep_i'}|\,,
$$
where we used Dirac's notations for states and linear forms on $\IC^D$. 
The norm of such an operator on $\IC^D$ is $|\cF^*_{\ep_i\ep'_i}|=D^{-1/2}$.
The norm of a tensor product operator is the product of the norms,
so for any $\bep$, $\bep'$ of length $n$, one has 
$\|U_{\bep, \bep'}\|= D^{-n/2}$. 
We thus get $c_n(U)=D^{-n/2}$, so that
\begin{equation}
h_{\cP^{(n)}}(\mu_k)\geq\frac{n}{2}\,\log D\,.
\end{equation}
This lower bound is slightly sharper than the one obtained in the previous paragraph,
Eq.~\eqref{e:lower}.  
However, the first approach seems more susceptible to generalizations, so we
decided to present it. The rest of the proof follows as before.\qed

\section{Lower bound on the topological entropy}\label{s:HTOP}

In this section, we prove the lower bound
for the topological entropies of supports of semiclassical measures (Theorem~\ref{HTOP}),
using the same strategy as for Anosov flows \cite{An}. Although, for the
case of the Walsh-baker,
this theorem is a consequence of Theorem \ref{HMET}, we decided to present this proof,
which does not use the Entropic Uncertainty Principle, but rather an
interplay of estimates between ``long'' logarithmic times, ``short'' logarithmic times and
finite times. As in the previous section, we are considering a certain 
sequence $\big(\psi_k\in\cH_{D^k}\big)$ of eigenstates of $B_k$, the 
Husimi measures of which converge to a semiclassical measure $\mu$, supported
on an invariant subset of $\t2$.

To prove Theorem~\ref{HTOP}, we consider an arbitrary closed invariant subset
$F\subset\t2$, which has a ``small'' topological entropy. Precisely,
we assume that 
$$
h_{top}(F)<\frac{\log D}{2}\,.
$$
Our aim is then to prove that
$\mu(F)<1$, implying that $F$ cannot be the support of $\mu$. 

\subsection{Finite-time covers of $F$}\label{s:finite-time}
The assumption on $h_{top}(F)$ implies that there exists
$\delta>0$, fixed from now on, such that
\begin{equation}\label{eq:assumption-F}
h_{top}(F)<\frac{\log D}{2}-10\,\delta\,.
\end{equation}
Given an integer $n_{o}$, we say that the set $W_{o}\subset \Sigma_{n_o}$
of $n_{o}$-cylinders
{\em covers} the set $F$ if and only if 
$$
F\subset \bigcup_{\bep\in W_o} [\cdot\bep]\,.
$$
In the limit of large lengths $n_o$, the topological entropy of $F$
measures the minimal cardinal of such covers. Precisely, let $N_{n_o}(F)$ be the
minimum cardinal for a set of $n_o$-cylinders covering $F$. For
the above $\delta>0$, there exists $n_{\delta}$ such that 
\begin{equation}\label{eq:topol-entropy}
\forall n_o\geq n_{\delta},\qquad N_{n_o}(F)\leq
\exp\big\{ n_o (h_{top}(F)+\delta )\big\}\,.
\end{equation}
Using the notations of Section~\ref{s:HMET}, 
the semiclassical measure of such a collection of $n_o$-cylinders is 
\begin{equation}\label{eq:nu(W_0)}
\mu(W_o)=\lim_{k\to\infty} \mu_k(W_{o})\,.
\end{equation}
On the other hand, from \eqref{e:sq-norm} we have, as long as $k\geq n_o$,
\begin{equation}\label{e:W_o}
\mu_k(W_o)=\Big\Vert \sum_{\bep\in W_{o}} P_{\bep}\,\psi_k\Big\Vert^2
=\sum_{\bep\in W_{o}}\la\psi_k,P_{\bep}\,\psi_k\ra\,.
\end{equation}
To show that $\mu(W_o)<1$, we would like to bound each term in the above sum.
Since the $P_{\bep}$ are orthogonal projectors, a trivial
bound for each term is $|\la\psi_k,P_{\bep}\,\psi_k\ra|\leq 1$. This is clearly not
sufficient for our aims. We therefore need a less direct method to bound
from above $\mu_k(W_o)$.

The next section presents the first step of this method. We show there that the 
norm of the operators $P_{\bep}$ satisfy 
exponential upper bounds for ``large logarithmic times'' $n$, namely when $n>k$
(we recall that $k=|\log h|/\log D$ is the Ehrenfest time of the system).

\subsection{Norms of the operators $P_{\bep}$}
The major ingredient in the proof of Theorem~\ref{HTOP} is an exponentially decaying
upper bound for
the norms of the operators $P_{\bep}$, for {\em arbitrarily large} times $n=|\bep|$. 
In the case of Anosov flows, such bounds require a heavy machinery \cite{An}.
In the present case, we are able to compute these norms exactly, in a rather straightforward
manner:
\begin{prop}\label{p:norms-B}
For any sequence $\bep$ of length $|\bep|=n$, the norm of 
the operator $P_{\bep}$ is given by
\begin{equation}\label{eq:estimate-B}
\| P_{\bep} \|=D^{-\max(0,n-k)/2}\,.
\end{equation}
\end{prop}
We see that the norm shows a ``transition'' at the Ehrenfest time $n=k$: it is
constant for $n\leq k$, and decreases exponentially for $n>k$. 

\begin{proof}
For $n\leq k$, $P_{\bep}$ is an orthogonal
projector, so the proposition is trivial in that case.

To deal with times $n>k$, we need to analyze the evolved projectors $P_{\ep}(l)$ 
coming into play in \eqref{e:B_bep} ($\ep=0,...,D-1$). Using \eqref{eq:action-observ} 
and the division 
$l=qk+r$, $r<k$, they can be written as:
$$
P_{\ep}(l)=(I)^{\otimes r}\otimes \cF^{q}\,\pi_{\ep}\,\cF^{-q}\otimes 
(I)^{\otimes k-r-1}\,.
$$
Hence, two evolved projectors $P_{\ep_1}(l_1)$, $P_{\ep_2}(l_2)$ will commute
with each other if $r_1\neq r_2$~: they act on different qu$D$its. As a result, 
within the product \eqref{e:B_bep}, we may group the factors $P_{\ep_l}(l-1)$
according to the equivalence class of $l$ modulo $k$, indexed by $r=0, \dots, k-1$. 
Each class contributes a product of $q'+1$ operators, of the form
\begin{gather}\label{e:class}
P_{\ep_{r+q'k+1}}(r+q'k)\cdots P_{\ep_{r+k+1}}(r+k)P_{\ep_r+1}(r)
=(I)^{\otimes r}\otimes A_{r+1}\otimes (I)^{\otimes k-r-1},\\
\text{where} \qquad
A_{r+1}=
\cF^{q'}\,\pi_{\ep_{r+1+q'k}}\,\cF^{-1}\,\pi_{\ep_{r+1+(q'-1)k}}\,\cF^{-1}\cdots 
\pi_{\ep_{r+1+k}}\,\cF^{-1}\,\pi_{\ep_{r+1}}\,.\nonumber
\end{gather}
Here $q'$ depends on $r$, it is the largest integer such that $r+1+q'k\leq n$.
Using Dirac's notations for states and linear forms
on $\IC^D$, the operator $A_{r+1}$ reads 
$$
A_{r+1}=\gamma_{r+1}\, \cF^{q'}|e_{\ep_{r+1+q'k}}\ra \la e_{\ep_{r+1}}|\,,
$$
where the prefactor $\gamma_{r+1}$ is the
product of $q'$ entries of the matrix $\cF^*$.
Since each entry has modulus $D^{-1/2}$, we obtain
$\| A_{r+1} \|=|\gamma_{r+1}|=D^{-q'/2}$.

There remains to count the number $q'+1$ of factors appearing in \eqref{e:class},
for each equivalence class
in the product~\ref{e:B_bep}. If we set $n=n_1k +n_2$, with $n_1\geq 1$ and
$n_2<k$, then each of 
the first $n_2$ classes (that is, such that $0\leq r\leq n_2-1$) contains $q'+1=n_1+1$ factors, 
while the remaining $k-n_2$ classes each contain $n_1$ factors. Since each equivalence class
acts on a different qu$D$it, the norm of $P_{\bep}$ is given
by
$$
\| P_{\bep}\|=\prod_{r=0}^{k-1} \| A_{r+1}\|= (D^{-n_1/2})^{n_2}
(D^{-(n_1-1)/2})^{k-n_2}=D^{(-n+k)/2}\,.
$$
\end{proof}
The estimate \eqref{eq:estimate-B} starts to be interesting
only for times $n>k$, that is beyond the Ehrenfest time.
On the other hand, the operators $P_{\bep}$ have a clear semiclassical meaning
(they project on the rectangles $[\cdot\bep]$) only when $n\leq k$. 
We need to connect these two disjoint time domains.

\subsection{Connecting ``long'' and ``short'' logarithmic times}
In this section we connect ``short logarithmic'' times $n\approx c k$,
$0<c\leq 1$
to ``long logarithmic'' times $n\approx C k$, with $C$ constant but arbitrary large.
To this aim, we fix $\theta\in (0, 1)$ and consider, for any $n\in\IN$, the sets 
$W_{n}\subset\Sigma_n$ of $n$-cylinders satisfying the following
condition:
\begin{equation}\label{eq:ineq1}
\Big\Vert \sum_{\bep\in \Sigma_n\setminus W_n} P_{\bep}\,\psi_k\Big\Vert \leq\theta\,.
\end{equation}
Such a set is called a $(k,1-\theta,n)$-cover of the state $\psi_k$. Intuitively, the
inequality
\eqref{eq:ineq1} means that the complement of $W_n$ in $\Sigma_n$, denoted by 
$\complement W_n$
in the sequel, has a small measure for the state $\psi_k$. 
We call $N_k(n,\theta)$
the minimal cardinal of a $(k,1-\theta,n)$-cover. 
Using the estimate \eqref{eq:estimate-B}, we can easily bound from
below this cardinal for ``large times'':
\begin{lem}\label{l:large-times} 
For any time $n>k$, the minimal cardinal of a $(k,1-\theta,n)$-cover
satisfies
\begin{equation}\label{e:large-times} 
N_k(n,\theta) \geq  D^{(n-k)/2}\,(1-\theta)\,.
\end{equation}
\end{lem}
Notice that the above lemma does not use the fact that $\psi_k$ is an eigenstate of $B_k$.

The next lemma is the crucial ingredient to connect the ``long times'' described by the lower bound
\eqref{e:large-times},
to the shorter times $n\approx ck$ ($0<c\leq 1$). 
This lemma uses the fact that $\psi_k$ is an eigenstate of
$B_k$.
\begin{lem}[Submultiplicativity]
\label{lem:(sub-multiplicativity)}
For any $1\leq n\leq k$, $1\leq \elll$ and $0<\theta<1$,
$$
N_k(\elll n,\theta)\leq N_k(n,\theta/\elll)^{\elll}\,.
$$
\end{lem}
\begin{proof}
Assume $W=W_{n}$ is a set satisfying \eqref{eq:ineq1} with $\theta/\elll$ instead of $\theta$.
Define $W^{\elll}$ as the set of sequences of length $n\elll$, formed of $\elll$ blocks
of length $n$,
$\bep^{(1)}\,\bep^{(2)}\,\ldots\,\bep^{(\elll)}$,
with all $\bep^{(i)}\in W$. Obviously, $\sharp (W^\elll)=(\sharp W)^\elll$.
To prove the lemma, it suffices to show that $W^{\elll}$ satisfies (\ref{eq:ineq1}). To do so,
we decompose the set $\complement(W^{\elll})=\Sigma_{\elll n}\setminus W^{\elll}$ in the
disjoint union:
\begin{equation}\label{eq:decompo}
\complement(W^{\elll})=\bigsqcup_{j=0}^{\elll-1}
\underbrace{\Sigma_{n}\cdots\Sigma_{n}}_{j}\;\complement W\;\underbrace{W\cdots W}_{\elll-j-1}.
\end{equation}
In other words, for a sequence of length $n\elll$ to belong
to the complement $\complement(W^{\elll})$, there must exist $0\leq j\leq \elll-1$ such that
the $j+1$-th block of length $n$ does not belong to $W$, the $j$ first blocks are arbitrary
and the $\elll-j-1$ last ones are in $W$.

In the sum $ \sum_{\bep\in \complement(W^{\elll})} P_{\bep}\,\psi_k$, each term
in the union \eqref{eq:decompo}
contributes 
\begin{multline*}
\Big(\sum_{\bep\in W} B_k^{-(\elll-1)n} P_{\bep} B_k^{(\elll-1)n}\Big) \cdots
\Big(\sum_{\bep\in W} B_k^{-(j+1)n} P_{\bep} B_k^{(j+1)n}\Big)
\Big(\sum_{\bep\in\complement W} B_k^{-jn} P_{\bep} B_k^{jn}\Big)\\
\times\Big(\sum_{\bep\in\Sigma_{n}} B_k^{(1-j)n} P_{\bep} B_k^{(j-1)n}\Big)\cdots
\Big(\sum_{\bep\in\Sigma_{n}} B_k^{-n} P_{\bep} B_k^{n}\Big)
\Big(\sum_{\bep\in\Sigma_{n}} P_{\bep}\Big)\psi_k\,.
\end{multline*}
Each sum on the second line yields the identity operator.
Because $\psi_k$ is an eigenstate of $B$, and using
the assumption on $W$, applying the last sum in the first line to $\psi_k$ gives a 
state of norm: 
$$
\Big\Vert \sum_{\bep\in\complement W} B_k^{-jn}P_{\bep} B_k^{jn} \,\psi_k \Big\Vert 
=\Big\Vert \sum_{\bep\in\complement W} P_{\bep}\, \psi_k \Big\Vert 
\leq\theta/\elll\,.
$$
Finally, from the fact that the $P_{\bep}$ are orthogonal
projectors for $|\bep|=n\leq k$, the previous sums in the first line are contracting operators:
$\Big\Vert \sum_{\bep\in W} P_{\bep}\Big\Vert \leq 1$.
As a result, each term of the union
\eqref{eq:decompo} corresponds to a state of norm $\leq\theta/\elll$. 
Finally summing over $j$, the triangle inequality leads to
$\Big\Vert 
\sum_{\bep\in\complement(W^{\elll})} P_{\bep}\, \psi_k \Big\Vert \leq\theta$.\end{proof}

Taking $n\approx c k$, $0<c\leq 1$ and $\elll>1/c$, we can now exploit 
both lemmas~\ref{l:large-times} and \ref{lem:(sub-multiplicativity)},
to get a lower
bound for the cardinals of $(k,1-\theta/\elll,n)$-covers:
$$
N_{k}(n,\theta/\elll) \geq D^{\frac{\elll n-k}{2\elll}}\;(1-\theta)^{1/\elll}\,.
$$
Taking $\elll>(c\delta)^{-1}$, and $\elll$ large enough so 
that $(1-\theta)^{1/\elll}>1/2$, this can be recast in the form:
\begin{equation}\label{eq:time-ck}
N_{k}(n,\theta/\elll)\geq \frac{1}{2}\,\exp\Big(\frac{n\log D}{2}(1-\delta)\Big)\,.
\end{equation}
This lower bound shows that a $(k,1-\theta/\elll,n)$-cover of the state $\psi_k$ cannot
be ``too thin''. 

\subsection{Connecting ``short logarithmic'' to finite times}
We need to use another trick to
relate the time $n\approx c k$, $0<c\leq 1$, to the fixed time $n_o$ considered in 
section~\ref{s:finite-time}.
This will finally yield an upper bound for $\mu_k(W_o)$, 
where $W_{o}$ is the union of $n_{o}$-cylinders covering $F$ described in
Section~\ref{s:finite-time}. 

The trick consists in using the following
sets of $n$-cylinders, defined relatively to $W_{o}$, and depending on
a parameter $\rho\in(0,1)$:
$$
\Sigma_{n}(W_{o},\rho)\defi\left\{ \bep\in\Sigma_{n}\: : \:
\frac{\sharp\left\{ 0\leq j\leq n-n_{o},\;
\ep_{j+1}\cdots\ep_{j+n_{o}}\in W_{o}\right\} }{n-n_{o}+1}
\geq\rho\right\}\, .
$$
This set is made of $n$-cylinders which will spend a fraction of
time larger than $\rho$ inside $W_{o}$, when evolved by the classical map.
A purely combinatorial argument (which we won't reproduce)
yields the following lemma:
\begin{lem}
Taking any $0<\rho<1$, $W_{o}\subset\Sigma_{n_o}$ fixed and $n>n_o$,
the cardinal of $\Sigma_{n}(W_{o},\rho)$ is bounded from above by
$$
\sharp\Sigma_{n}(W_{o},\rho)\leq\binom{\lfloor n/n_{o}\rfloor}{n}^{2}\times
\left(\sharp W_{o}\right)^{[n/n_{o}]}\times D^{(1-\rho)n_{o}n+n_{o}}.
$$
\end{lem}
Let us take $n_o$ large enough such that, in the limit $n\to\infty$,
the first binomial factor is less than $\e^{\delta n}$. Then, take for
$W_{o}\in\Sigma_{n_o}$ a cover of $F$, with its cardinal bounded from above by 
\eqref{eq:topol-entropy}. 
For $n$ large enough, the above upper bound then becomes~:
\begin{equation}\label{eq:ineq-mu}
\sharp\Sigma_{n}(W_{o},\rho)\leq\e^{2\delta n}\;\e^{n\left(h_{top}(F)+\delta\right)}\;
\e^{\{(1-\rho)n_{o}n+n_o\}\log D}\leq
\e^{n\{h_{top}(F)+(1-\rho) n_{o}\log D+4\delta\}}\,.
\end{equation}
Let us take $\rho$ sufficiently close to $1$, such that $(1-\rho)n_{o}\log D+4\delta\leq 5\delta$.
In that case, comparing the growth rate with \eqref{eq:time-ck} and
the assumption \eqref{eq:assumption-F} on $h_{top}(F)$, we see that
the sets $\Sigma_{n}(W_{o},\rho)$ are too small to cover $\psi_k$: 
$$
\Big\Vert \sum_{\bep\in\complement\Sigma_{n}(W_{o},\rho)}P_{\bep}\,\psi_k\Big\Vert 
\geq\theta/\elll.
$$
Because the operators $P_{\bep}$ are orthogonal
projectors, this inequality can be written
$$
\mu_k\big(\complement\Sigma_{n}(W_{o},\rho)\big)=
\sum_{\bep\in\complement\Sigma_{n}(W_{o},\rho)}\la P_{\bep}\,\psi_k, \psi_k\ra=
\Big\Vert \sum_{\bep\in\complement\Sigma_{n}(W_{o},\rho)}P_{\bep}\,\psi_k\Big\Vert^2 
\geq(\theta/\elll)^2, 
$$
so that
\begin{equation}\label{eq:rho_h(Sigma)}
\mu_k\big(\Sigma_{n}(W_{o},\rho)\big)
\leq 1- (\theta/\elll)^2\,.
\end{equation}
We are now ready to compute $\mu_k(W_{o})$: 
\begin{equation}\label{e:sum-Wo}
\begin{split}
\mu_k(W_{o})&=\la\psi_k,\,\Big(\sum_{\bep\in W_{o}}
\frac{1}{n-n_{o}+1}\sum_{j=0}^{n-n_{o}}B^{-j}P_{\bep} B^{j} \Big)\,\psi_k\ra\,\\
&= \sum_{\bal\in \Sigma_{n}} \la \psi_k,\,P_{\bal}\,\psi_k \ra
\left(\frac{\sharp\left\{ 0\leq j\leq n-n_{o},\;
\alpha_{j+1}\cdots\alpha_{j+n_{o}}\in W_{o}\right\} }{n-n_{o}+1}\right)
\end{split}
\end{equation}
In the first line, we used the fact that $\psi_k$ is an eigenstate of $B_k$.
To get the second line, we have written $B^{-j}P_{\bep} B^{j}$ as
$$
B^{-j}P_{\bep} B^{j}=\sum_{\bal \in \Sigma_{n}, \alpha_{j+1}=\ep_1,\cdots, \alpha_{j+n_o}=
\ep_{n_o}} P_{\bal}\,,
$$
and rearranged the sum.

By definition, an $n$-cylinder $[.\bal]$ belongs
to $\Sigma_{n}(W_{o},\rho)$ if and only if
its corresponding coefficient 
$\frac{\sharp\left\{ 0\leq j\leq n-n_{o},\;
\alpha_{j+1}\cdots\alpha_{j+n_{o}}\in W_{o}\right\} }{n-n_{o}+1}$ is greater than $\rho$. 
As a consequence, \eqref{e:sum-Wo} is bounded from above by
$$
\mu_k(W_{o})\leq\mu_k\big(\Sigma_{n}(W_{o},\rho)\big)
+\rho\,\mu_k\big(\complement\Sigma_{n}(W_{o},\rho)\big)\,.
$$
Using the upper bound \eqref{eq:rho_h(Sigma)} for the measure of $\Sigma_{n}(W_{o},\rho)$, 
we obtain
$$
\mu_k(W_{o})\leq(1-\rho)\left(1-\left(\theta/\elll\right)^{2}\right)+\rho.
$$
Finally, we may send $k\to\infty$, and use \eqref{eq:nu(W_0)} to get the required upper
bound:
$$
\mu(F)\leq\mu(W_{o})\leq(1-\rho)\left(1-\left(\theta/\elll\right)^{2}\right)+\rho<1\,.
$$
This ends the proof of Theorem~\ref{HTOP}.

\section{Proof of Theorem \ref{t:loc-erg}}\label{s:loc-erg}
Since the proof of the theorem is the same as 
for the cat map \cite{FN04}, we will only explain the strategy
for a sequence of eigenstates $(\psi_k)$
converging towards an invariant measure $\mu$ of the following form:
\begin{equation}\label{e:mu}
\mu=\beta\delta_{(0)}+(1-\beta)\nu\,,
\end{equation}
where $\delta_{(0)}$ is the delta measure on
the fixed point $(0)\defeq\ldots000\cdot000\ldots$ of $\Sigma$
(which maps to the origin of the torus), and $\nu$ is any 
invariant probability measure
on $\Sigma$ which does not charge $(0)$. We will prove the
\begin{prop}
A semiclassical measure $\mu$ of the form \eqref{e:mu} necessarily 
contains a Lebesgue component of weight larger or equal to $\beta$.
\end{prop}
The same statement holds (with a similar proof) if we replace $\delta_{(0)}$ by a finite
combination of Dirac measures on periodic orbits, and directly gives
Theorem~\ref{t:loc-erg}.
\begin{proof}
To localize on $(0)$, we will consider the rectangles
$R_\ell\defeq [0_\ell\cdot 0_\ell]$, where $0_\ell$ 
is the sequence of length $\ell$ only made of zeros. As long as $\ell\leq \lfloor k/2 \rfloor$, 
the characteristic function on $R_{\ell}$ is quantized into an orthogonal projector:
$$
\Op_{k,\lfloor k/2 \rfloor}(\bbbone_{R_\ell})=(\pi_0)^{\otimes \ell}\otimes(I)^{\otimes k-2\ell}
\otimes(\cF_D^*\pi_0\cF_D)^{\otimes \ell}\defeq P_{R_\ell}\,.
$$

Because the sequence of eigenstates $(\psi_k)$ converges 
towards $\mu$, it is possible to 
find a sequence $\ell(k)\to\infty$ such that
\begin{equation}\label{e:delta0}
\la\psi_k,\Op_{k,\lfloor k/2 \rfloor}(\bbbone_{R_{\ell(k)}})\,\psi_k\ra
\stackrel{k\to\infty}{\longrightarrow} \beta\,.
\end{equation}
The divergence of the sequence $\ell(k)$ can be taken arbitrarily slow, so we
can assume that $\ell(k)<k/2$ for all $k$.
Equipped with such a sequence, we decompose $\psi_k$ into 
$\psi_k=\psi_{k,(0)}+\psi_{k,\nu}$ with
$$
\psi_{k,(0)}\defeq P_{R_{\ell(k)}}\,\psi_k,\qquad
\psi_{k,\nu}\defeq \big(1-P_{R_{\ell(k)}}\big)\,\psi_k\,.
$$
Equation \eqref{e:delta0}, together with the assumptions on $\mu$, show that 
the Walsh-Husimi measures of $\psi_{k,(0)}$, resp. $\psi_{k,\nu}$, 
converge to the measure $\beta\delta_{(0)}$, resp. $(1-\beta)\nu$.

The observables we will use to test the various measures
are characteristic functions on rectangles $R=[\bep'\cdot\bep]$ of lengths $n'+n$. 
For $k$ large enough, such a fixed rectangle is quantized into the orthogonal
projector
$$
P_{R}=
\pi_{\ep_1}\otimes\ldots\otimes \pi_{\ep_n}\otimes I\otimes\ldots I\otimes 
\cF^*\pi_{\ep'_{n'}}\cF
\otimes\ldots\otimes\cF^*\pi_{\ep'_{1}}\cF\,.
$$
To prove the theorem, we will consider the matrix elements
$\la\psi_k,\,P_{R}\,\psi_k\ra$, which by assumption converges to
$\mu(R)$ as $k\to \infty$.

Since $\psi_k$ is an eigenstate of $B_k$, we can replace $P_R$ by 
$P_R'\defeq {B_k}^{-k}\,P_{R}\,{B_k}^k$ in this matrix element, and then
split the eigenstate:
\begin{equation}\label{e:decompo}
\la\psi_k,\,P_{R}\,\psi_k\ra=
\la\psi_{k,(0)},\,P_{R}'\,\psi_{k,(0)}\ra+
\la\psi_{k,\nu},\,P_{R}'\,\psi_{k,\nu}\ra+
2\Re\big(\la\psi_{k,(0)},\,P_{R}'\,\psi_{k,\nu}\ra\big)\,.
\end{equation}
Using \eqref{e:power-k}, we easily compute $P_{R}'$:
$$
P_{R}'=
\cF\pi_{\ep_1}\cF^*\otimes\ldots\otimes \cF\pi_{\ep_n}\cF^*
\otimes I\otimes\ldots I\otimes \pi_{\ep'_{n'}}\otimes\ldots\otimes\pi_{\ep'_{1}}\,.
$$
In the first term on the right hand side of \eqref{e:decompo}, 
this operator is sandwiched between two projectors $P_{R_{\ell(k)}}$. By taking $k$ large
enough, we make sure that $\ell=\ell(k)\geq\max(n,n')$. Under this condition,
$P_{R_{\ell}}\,P_{R}'\,P_{R_{\ell}}$ is a tensor product operator, with
each of the $n$ first tensor factors of the form 
$$
\pi_0\cF\pi_{\ep_i}\cF^*\pi_0=|\cF_{0\ep_i}|^2\,\pi_0=D^{-1}\,\pi_0\,.
$$
Similarly, each of its $n'$ last factors reads
$D^{-1}\,\cF^*\pi_0\cF$, while the remaining $k-n-n'$ factors
inbetween make up
\begin{equation}\label{e:A-center}
A_{center}\defeq (\pi_0)^{\otimes(\ell-n)}\otimes 
(I)^{\otimes(k-2\ell)}\otimes
(\cF^*\pi_0\cF)^{\otimes(\ell-n')}\,.
\end{equation}
As a result, 
$P_{R_{\ell}}\,P_{R}'\,P_{R_{\ell}}=D^{-n-n'}\,P_{R_{\ell}}$.
From the definition of $\psi_{k,(0)}$, this implies that 
\begin{equation}\label{e:1st-term}
\lim_{k\to\infty}\la\psi_{k,(0)},\,P_{R}'\,\psi_{k,(0)}\ra
=\beta\,D^{-n-n'}=\beta\, \mu_{Leb}(R)\,.
\end{equation}
This identity shows that the states ${B_k}^k\,\psi_{k,(0)}$ are semiclassically
equidistributed, as in the case of the cat map \cite[Prop.~3.1]{FN04}.
Due to the positivity of the operator $P_{R}'$,
the second term on the right hand side of \eqref{e:decompo} is positive.

The last term in \eqref{e:decompo} is dealt with in the following lemma, analogous
to \cite[Prop.~3.2]{FN04}:
\begin{lem}\label{l:0-nu}
With the above notations, we have
$$
\lim_{k\to\infty}\la\psi_{k,(0)},\,P_{R}'\,\psi_{k,\nu}\ra=0\,.
$$
\end{lem}
With this lemma, \eqref{e:decompo} and \eqref{e:1st-term}, we deduce that 
$$
\mu(R)=\lim_{k\to\infty} \la\psi_{k},\,P_{R}\,\psi_{k}\ra=
\lim_{k\to\infty} \la\psi_{k},\,P_{R}'\,\psi_{k}\ra \geq 
\lim_{k\to\infty}\la\psi_{k,(0)},\,P_{R}'\,\psi_{k,(0)}\ra=
\beta\, \mu_{Leb}(R).
$$
This shows that the Lebesgue component of $\mu$ 
necessarily has a weight $ \geq\beta$.
\end{proof}

\subsubsection*{Proof of Lemma \ref{l:0-nu}}
We want to prove that $\la\psi_k,\,P_{R_\ell}P_{R}'(1-P_{R_\ell})\,\psi_k\ra$ vanishes
as $k\to\infty$. We start by expanding the operator
$P_{R_{\ell}}P_{R}'$. Its first $n$ tensor factors are of the type
\begin{equation}\label{e:factor1}
\pi_0\cF\pi_{\ep_i}\cF^*=D^{-1}\,\pi_0+\sum_{\alpha=1}^{D-1}
\cF_{0\ep_i}\cF^*_{\ep_i\alpha}\,|e_0\ra \la e_{\alpha}|
\end{equation}
The subsequent $k-n-n'$ factors make up the operator $A_{center}$ described above,
and the last $n'$ factors have the form
\begin{equation}\label{e:factor2}
\cF^*\pi_0\cF\pi_{\ep'_i}=D^{-1}\,\cF^*\pi_0\cF+
\sum_{\alpha=1}^{D-1}\cF_{0\ep'_i}\cF^*_{\ep'_i \alpha}\ 
\cF^*|e_0\ra \la e_{\alpha}|\cF\,.
\end{equation}
In (\ref{e:factor1},\ref{e:factor2}) we voluntarily separated from the sum the 
term appearing in the tensor decomposition of $D^{-n-n'}\,P_{R_{\ell}}$.
As a consequence, the operator $P_{R_{\ell}}P_{R}'$ can be written
as the sum of $D^{n+n'}$ operators of the form
\begin{equation}\label{term1}
D^{-n-n'}\,A_{\bal}\otimes A_{center}\otimes A'_{\bal'}\,,
\end{equation}
where we use \eqref{e:A-center} and the tensor products
$$
A_{\bal}=\e^{\i \varphi(\bep,\bal)}\;
(|e_0\ra\la e_{\alpha_1}|)\otimes \ldots\otimes(|e_0\ra\la e_{\alpha_n}|),\quad
A'_{\bal'}=\e^{\i \varphi'(\bep',\bal')}\;
(\cF^*|e_0\ra\la e_{\alpha'_{n'}}|\cF)\otimes 
\ldots\otimes(\cF^*|e_0\ra\la e_{\alpha'_1}|\cF)\,.
$$
The phase prefactors are not important, so we omit their explicit expression.
The sequences
$\bal'\cdot\bal=\alpha'_{n'}\ldots\alpha'_1\cdot\alpha_1\ldots\alpha_n$ can
take all values in $(\IZ_D)^{n'+n}$. 

The term 
$A_{0_n}\otimes A_{center}\otimes A'_{0_{n'}}$ exactly equals
the projector $P_{R_\ell}$, so that $P_{R_{\ell}}P_{R}'(1-P_{R_{\ell}})$
is the sum of the terms \eqref{term1} over all sequences $\bal'\cdot\bal\neq 0_{n'}\cdot 0_{n}$.
Our last task consists in proving that for any such sequence, 
\begin{equation}\label{e:last-step}
\la\psi_k,\,A_{\bal}\otimes A_{center}\otimes A'_{\bal'}\,\psi_k\ra
\stackrel{k \to\infty}{\longrightarrow}0\,.
\end{equation}
From the structure of $A_{\bal}$ and $A'_{\bal'}$, 
this scalar product is unchanged if we replace the state $\psi_k$ on
the right by its projection on the rectangle 
$\tilde{R}_{\ell}=[0_{\ell-n'}\bal'\cdot\bal 0_{\ell-n}]$. Because the above operator
has norm unity and $\psi_k$ is normalized, 
the left-hand side of \eqref{e:last-step} is bounded from above by 
$\norm{P_{\tilde R_{\ell}}\,\psi_k}$. For any $m\geq \max(n,n')$, the 
rectangle $\tilde R_{\ell}$ is contained in 
$\tilde R_{m}=[0_{m-n'}\bal'\cdot \bal 0_{m-n}]$ as soon
as $\ell=\ell(k)\geq m$, so that 
$\norm{P_{\tilde R_{\ell(k)}}\,\psi_k}\leq \norm{P_{\tilde R_{m}}\,\psi_k}$.
On the other hand, we know that
$\norm{P_{\tilde R_{m}}\,\psi_k}^2$ converges to $\mu(\tilde R_{m})$ as
$k\to\infty$.

We finally use the fact that $\mu$ is an invariant probability measure
to show that 
$\mu(\tilde R_{m})\stackrel{m\to\infty}{\longrightarrow}\nolinebreak 0$.
Indeed, in this limit, the rectangles $\tilde R_{m}$ shrink to the point
$\ldots 00\bal'\cdot\bal00\ldots$, which is homoclinic to the fixed point $(0)$. 
If $\mu$ were charging that point, it would equally charge all its iterates, which
form an infinite orbit: this would violate the normalization of $\mu$. 
Finally, we can find a sequence $m(k)\to\infty$ such 
that $m(k)\leq\ell(k)$ and $\norm{P_{\tilde R_{m(k)}}\,\psi_k}\to 0$, which
proves \eqref{e:last-step}.
The lemma follows by summing over the finitely many 
sequences $\bal'\cdot\bal$ of length $n'+n$.
\qed

\appendix

\section{The Entropic Uncertainty Principle}

Let us recall the statement of the Riesz interpolation theorem (also called ``Riesz convexity
theorem''), in the basic
case when it is applied to
a linear operator $T$ acting on $\IC^{N}$. We denote $l_p(N)$ the Banach space
obtained by endowing $\IC^{N}$ with the norm
$$
\Vert\psi\Vert_p=\left(\sum_{j=1}^N |\psi_j|^p \right)^{1/p}, 
$$
where $\left(\psi_{j}\right)_{j=1,\ldots, N}$ is the representation of $\psi$ 
in the canonical basis.
We also denote
$$
\Vert\psi\Vert_\infty=\max\{|\psi_j|, j=1,\ldots, N\}.
$$
We are interested in the norm $\left\Vert T\right\Vert_{p,q}$ 
of the operator $T$, acting from
$l_{p}$ to $l_{q}$, for $1\leq p,q\leq\infty$. The following
theorem holds true \cite[Section VI.10]{DunSchw}:
\begin{thm} [Riesz interpolation theorem]
The function $\log\left\Vert T\right\Vert _{1/a,1/b}$ is a convex
function of $(a,b)$ in the square $0\leq a,b\leq1$.
\end{thm}
From this theorem, we now reproduce the derivation of Maassen and Uffink \cite{MaaUff}
to obtain nonstandard uncertainty relations.
We denote $(T_{jk})$ the matrix
of $T$ in the canonical basis.
In the case
$a=1,b=0$, we have for any $\psi$
$$
\left\Vert T\psi\right\Vert_{\infty}=
\sup_{j}\left|(T\psi)_{j}\right|\leq
\sup_{j,k}\left|T_{j,k}\right|\,\sum_{k'}\left|\psi_{k'}\right|=
\sup_{j,k}\left|T_{j,k}\right|\,\left\Vert \psi\right\Vert_{1}\,,
$$
which can be written as 
$\left\Vert T\right\Vert _{1,\infty}\leq\sup_{j,k}\left|T_{j,k}\right|\defi c(T)$.

Let us assume that $T$ is contracting on $l_2$~: $\left\Vert T\right\Vert_{2,2}\leq1$.
We take $t\in[0, 1]$ and  $a_{t}=\frac{1+t}{2}$, $b_{t}=\frac{1-t}{2}$ to interpolate between
$(1/2,1/2)$ and $(1,0)$; the above
theorem implies that 
$$
\left\Vert T\right\Vert _{1/a_{t},1/b_{t}}\leq c(T)^{t}\,.
$$
This is equivalent to the following
\begin{cor}
\label{pro:Maassen-Uffink}Let the $N\times N$ matrix $T$ satisfy
$\left\Vert T\right\Vert _{2,2}\leq1$ and call $c(T)\defi\sup_{j,k}\left|T_{j,k}\right|$.
Then, for all $t\in[0,1]$, for all $\psi\in\IC^{N}$,
$$
\left\Vert T\psi\right\Vert_{\frac{2}{1-t}}\leq 
c(T)^{t}\,\left\Vert \psi\right\Vert_{\frac{2}{1+t}}\,.
$$
\end{cor}
Keeping the
notations of \cite{MaaUff}, 
we define for any $r>0$ and $-1<r<0$ the ``moments''
$$
M_{r}(\psi)\defi\left(\sum_{j}\left|\psi_{j}\right|^{2+2r}\right)^{1/r}\,.
$$
The above corollary leads to the following family of 
``uncertainty relations'':
\begin{equation}\label{eq:ineq2}
\forall t\in(0,1),\ \forall\psi\in\IC^D, \qquad M_{\frac{t}{1-t}}(T\psi)\, 
M_{\frac{-t}{1+t}}(\psi)\leq c(T)^{2}\,.
\end{equation}
In the case $\|\psi\|_{2}=1$, we notice that the moments
converge to the same value when $r\to 0$ from above or below:
$$
\lim_{r\to 0} M_{r}(\psi)=\e^{-h(\psi)}\,,\quad\text{where}\quad
h(\psi)=-\sum_{j}\left|\psi_{j}\right|^{2}\log\left|\psi_{j}\right|^{2}\,.
$$
If furthermore $\|T\psi\|_2=1$, in particular if $T=U$ is unitary, 
then the limit $t\to 0$ of the inequalities 
\eqref{eq:ineq2} yield the Entropic Uncertainty Principle stated in 
Theorem~\ref{thm:entropic_uncertainty}.

\subsection*{Vectorial Entropic Uncertainty Principle}
This theorem can be straightforwardly generalized in the following way. 
Let $(\cH, \Vert.\Vert)$ be a Hilbert space, and suppose
we are given a family of orthogonal projectors $(P_j)_{j=1,\ldots, N}$ on $\cH$, satisfying
\begin{equation}\label{e:partition}
P_i P_j=\delta_{ij}\,P_i,\qquad \sum_{j=1}^N P_j=I_{\cH}\,.
\end{equation}
Using these operators, we decompose any $\Psi\in\cH$ into the states
$\Psi_j\defeq P_j\Psi$. The above identity 
implies that
$$
\Vert\Psi\Vert^2=\sum_j \Vert \Psi_j\Vert^2\,.
$$
Using this decomposition, the vector space $\cH$ 
can be endowed with different norms, all equivalent
to the Hilbert norm $\Vert.\Vert$ since $N$ is finite~:
$$
\Vert\Psi\Vert_p\defi \left(\sum_{j=1}^N \Vert\Psi_j\Vert^p \right)^{1/p}, \qquad
\Vert\Psi\Vert_\infty=\max\{\Vert\Psi_j\Vert, j=1,\ldots, N\}\,.
$$
Notice that $\|\Psi\|_2=\|\Psi\|$.

Given a bounded operator $T$ on $\cH$, we define the operators 
$T_{jk}=P_j T P_k$, in terms of which $T$
acts on $\Psi\in\cH$ as follows:
$$
(T\Psi)_j=\sum_{k}T_{jk}\Psi_k.
$$
Let us denote $c(T)=\max_{j,k} \left\Vert T_{jk}\right\Vert$.
The Riesz interpolation theorem still holds in this setting, and yields, provided
$\Vert T\Vert=\Vert  T\Vert_{2,2}\leq 1$, 
\begin{equation}
\forall t\in [0,1],\ \forall \Psi\in\cH,\qquad 
\left\Vert T\Psi\right\Vert_{\frac{2}{1-t}}\leq 
c(T)^{t}\,\left\Vert \Psi\right\Vert _{\frac{2}{1+t}}\,.
\end{equation}
This implies the following vectorial Entropic Uncertainty Principle, which we
use in Section~\ref{ss:GEUP}~:
\begin{thm}\label{a:EUP}
Let $U$ be a unitary operator on $\cH$, and, using a 
partition of unity \eqref{e:partition}, define 
$c(U)\defi\max_{j,k}\left\Vert U_{jk}\right\Vert$ and, 
for any normalized $\Psi\in\cH$, the entropy
$$
h(\Psi)=-\sum_{j}\left\Vert\Psi_{j}\right\Vert^{2}\log\left\Vert\Psi_{j}\right\Vert^{2}\,.
$$
This entropy satisfies the following inequality:
$$
h(U\Psi)+h(\Psi)\geq-2\log c(U)\,.
$$
\end{thm}


\end{document}